\documentclass[12pt,onecolumn,journal]{IEEEtran}
\usepackage{amsmath,amssymb,amsthm}
\usepackage{algorithm}
\usepackage{algpseudocode}
\usepackage{graphicx}
\usepackage[colorlinks=true,linkcolor=blue,citecolor=blue,urlcolor=blue]{hyperref}
\usepackage{xcolor}
\usepackage{placeins}
\newtheorem{theorem}{Theorem}
\newtheorem{proposition}{Proposition}
\newtheorem{lemma}{Lemma}
\newtheorem{corollary}{Corollary}
\newtheorem{definition}{Definition}
\newtheorem{assumption}{Assumption}
\newtheorem{remark}{Remark}

\newcommand{\R}{\mathbb{R}}
\newcommand{\Sym}{\mathbb{S}}
\newcommand{\X}{\mathcal{X}}
\newcommand{\D}{\mathcal{D}}
\newcommand{\lmax}{\lambda_{\max}}
\providecommand{\tr}{}
\renewcommand{\tr}{\operatorname{tr}}
\IfFileExists{numbers.tex}{

\providecommand{\PolyRoundCap}{12}
\providecommand{\PolyTrainSteps}{500}
\providecommand{\PolyLearningRate}{0.02}
\providecommand{\PolyInitialSamples}{500}
\providecommand{\OutcomeGrowthFactor}{100}
\providecommand{\numSeeds}{20}
\providecommand{\numGridCoarse}{121}
\providecommand{\numGridOp}{2001}
\providecommand{\numTauOp}{0.00142}
\providecommand{\numEpsZero}{0.05}
\providecommand{\numGamma}{1.3}
\providecommand{\numStallCount}{19}

\providecommand{\numDivCount}{20}
\providecommand{\numDivGrowth}{9.9 \times 10^{6}}
\providecommand{\numCertCount}{19}
\providecommand{\numCertMedIters}{2}
\providecommand{\numGridMedianRatio}{0.0057}
\providecommand{\numGridMinRatio}{0.00164}
\providecommand{\numGridMaxRatio}{0.0145}
\providecommand{\numGridB}{4.25}
\providecommand{\numGridPtsTwo}{2.5 \times 10^{5}}
\providecommand{\numGridPtsFour}{2.4 \times 10^{11}}
\providecommand{\numGridPtsSix}{4.1 \times 10^{17}}
\providecommand{\numEta}{0.0386}
\providecommand{\numBupper}{5.94}
\providecommand{\numMuOne}{1.87}
\providecommand{\numCdZero}{1.42}
\providecommand{\numCdOne}{1.23}
\providecommand{\numDeltaStar}{0.00224}
\providecommand{\numDeltaEmpRatio}{4}
\providecommand{\PendRoundCap}{10}
\providecommand{\PendTrainSteps}{500}
\providecommand{\PendLearningRate}{0.02}
\providecommand{\PendInitialSamples}{500}
\providecommand{\numPendSeeds}{20}
\providecommand{\numPendCert}{19}
\providecommand{\numPendDiv}{1}
\providecommand{\numPendMedIters}{1}
\providecommand{\numPendRate}{0.2}
\providecommand{\numPendBox}{1}
\providecommand{\numPendGrid}{401}
\providecommand{\numPendL}{1.56}
\providecommand{\numPendSlack}{0.0487}
\providecommand{\numPendB}{2.25}
\providecommand{\NeuralRoundCap}{8}
\providecommand{\NeuralTrainSteps}{600}
\providecommand{\NeuralLearningRate}{0.01}
\providecommand{\NeuralInitialSamples}{800}
\providecommand{\numNeuralUnits}{6}
\providecommand{\numNeuralSeeds}{20}
\providecommand{\numNeuralCert}{13}
\providecommand{\numNeuralCoarse}{1}
\providecommand{\numNeuralDiv}{7}
\providecommand{\numNeuralMedIters}{1}
\providecommand{\numNeuralOpGrid}{8001}
\providecommand{\numNeuralCoarseGrid}{2001}
\providecommand{\numNeuralL}{54.1}
\providecommand{\numNeuralSlack}{0.0281}
\providecommand{\numNeuralUniformSlack}{0.00908}
\providecommand{\numNeuralA}{0.1}
\providecommand{\numNeuralB}{5.24}

\providecommand{\numCertA}{0.1}
}{}
\IfFileExists{revision_numbers.tex}{
\providecommand{\revPerturbStd}{0.08}
\providecommand{\revPullbackMinEigLower}{0.0555}
\providecommand{\revPullbackThreshold}{4.21}
\providecommand{\revOtherPerturbStd}{0.05}
\providecommand{\revResidualDepth}{3}
\providecommand{\revResidualGridAxis}{4001}
\providecommand{\revResidualSeed}{0}
\providecommand{\revExtensionSettings}{The new benchmark uses $[-1,1]^3$, rate $0.4$, floor $0.1$ and $20$ seeds. Each retraining round has at most $300$ Adam steps, with $256$ initial samples and at most $8$ retraining rounds. The oracle uses at most $12000$ boxes and depth $24$ per call, with target raw margin $1 \times 10^{-4}$. The learner uses a fixed normalized hinge margin $0.005$ and learning rate $0.02$; it does not use the grid loop's coupled margin. Zero-loss retraining that leaves the parameters unchanged stops an inconclusive run.}
\providecommand{\revSurrogateSettings}{Quadratic regression uses $12$ noiseless trajectories with $25$ sampled times each; central differences use time step $0.01$. The basis contains every monomial of total degree at most two. Its metric learner uses a fixed normalized hinge margin $0.05$, $400$ initial samples, at most $400$ Adam steps per round, learning rate $0.02$ and at most $8$ retraining rounds; the same unchanged-parameter stopping rule applies.}
\providecommand{\revBoundedRadius}{The radius is computed before the runs as four times the Frobenius norm of the common initial factor coefficients; its stored value is $R\approx 5.65685$.}
\providecommand{\revSensitivitySettings}{The sensitivity diagnostic uses $5$ seeds, a common $31^2$ grid, and at most $4$ retraining rounds with $500$ Adam steps each.}
\providecommand{\revLegacySeedsSummary}{The fixed-margin coarse-grid exception, seed $8$, certified after $2$ retraining rounds. The coupled operating-grid exception, seed $15$, met the finite-budget divergence criterion after $12$ retraining rounds.}
\providecommand{\revAblationSummary}{Table~\ref{tab:ablation} reports every regime, including uncertified runs.}
\providecommand{\revScaleChaseSummary}{Figure~\ref{fig:scalechase} illustrates the constructed scaling sequence; it does not estimate optimizer failure frequencies.}
\providecommand{\revExtensionSummary}{Table~\ref{tab:extensions} reports the three-dimensional synthesis and matched-candidate oracle checks. The uncertified runs exhaust the box budget and stop after zero-loss retraining leaves the parameters unchanged.}
\providecommand{\revSurrogateSummary}{Table~\ref{tab:extensions} includes global residual bounds for the fitted surrogate and the transferred rate.}
}{}
\IfFileExists{application_numbers.tex}{
\providecommand{\appBLarge}{0.75}
\providecommand{\appBMedium}{0.5}
\providecommand{\appBSmall}{0.25}
\providecommand{\appBatch}{512}
\providecommand{\appBoxBudget}{8191}
\providecommand{\appCegisLarge}{0}
\providecommand{\appCegisMedium}{0}
\providecommand{\appCegisSmall}{20}
\providecommand{\appConstKLower}{0.27}
\providecommand{\appConstKUpper}{0.7}
\providecommand{\appDepthBudget}{24}
\providecommand{\appFixedLarge}{0}
\providecommand{\appFixedMedium}{0}
\providecommand{\appFixedSmall}{20}
\providecommand{\appFloor}{0.1}
\providecommand{\appLearningRate}{0.01}
\providecommand{\appParameters}{39}
\providecommand{\appPdiDiscordantSmall}{2}
\providecommand{\appPdiLarge}{0}
\providecommand{\appPdiMedium}{0}
\providecommand{\appPdiSmall}{18}
\providecommand{\appRate}{0.05}
\providecommand{\appRefCegisLarge}{0}
\providecommand{\appRefCegisMedium}{2}
\providecommand{\appRefCegisSmall}{20}
\providecommand{\appRefFixedLarge}{0}
\providecommand{\appRefFixedMedium}{0}
\providecommand{\appRefFixedSmall}{20}
\providecommand{\appRefPdiLarge}{0}
\providecommand{\appRefPdiMedium}{0}
\providecommand{\appRefPdiSmall}{18}
\providecommand{\appRefinementBudget}{65535}
\providecommand{\appReturned}{64}
\providecommand{\appRounds}{5}
\providecommand{\appSamples}{512}
\providecommand{\appScreen}{33}
\providecommand{\appSeeds}{20}
\providecommand{\appSteps}{400}
\providecommand{\appTotalRuns}{180}
\providecommand{\appTotalSteps}{2000}
\providecommand{\appTrainMargin}{0.01}
\providecommand{\appUnits}{6}
\providecommand{\appVerifyMargin}{1 \times 10^{-4}}
\providecommand{\appEnergySmall}{0.01196}
\providecommand{\appEnergyMedium}{0.03905}
\providecommand{\appEnergyLarge}{0.06745}
}{}
\providecommand{\polyConstRate}{\frac{2}{5}}
\providecommand{\polyConstPFirst}{12}
\providecommand{\polyConstPSecond}{1}
\providecommand{\polyConstDiagonalThreshold}{\frac{100}{9}}
\providecommand{\polyConstCFirst}{\frac{72}{5}}
\providecommand{\polyConstCSecond}{\frac{6}{5}}
\providecommand{\polyConstCross}{2}
\providecommand{\polyConstRawMargin}{\frac{1}{20}}
\providecommand{\polyConstNormalizedMargin}{\frac{1}{260}}
\providecommand{\polyConstBox}{2}
\providecommand{\polyConstMetricUpper}{12}
\providecommand{\polyConstShiftLeading}{\frac{287}{20}}
\providecommand{\polyConstShiftDet}{\frac{201}{400}}

\title{Neurosymbolic Synthesis of Contraction Metrics with Certified Verification}
\hypersetup{pdftitle={Neurosymbolic Synthesis of Contraction Metrics with Certified Verification},pdfauthor={Victor M. Preciado}}

\author{Victor~M.~Preciado,~\IEEEmembership{Senior~Member,~IEEE}%
\thanks{V.~M.~Preciado is with the Department of Electrical and Systems
Engineering and the Department of Computer and Information Science,
University of Pennsylvania, Philadelphia, PA 19104 USA
(e-mail: preciado@seas.upenn.edu).}}

\begin{document}
\maketitle

\begin{abstract}
We study counterexample-guided synthesis of contraction metrics for continuous-time autonomous systems on compact domains. A learner proposes a metric; a sound verifier certifies its matrix inequality or returns diagnostic points for retraining. The neural extension pairs a learned metric factor with symbolic contraction conditions and rigorous numerical verification; main ablations use polynomial features. We give computable polynomial bounds in arbitrary dimension and outward-rounded verification for polynomial, trigonometric, and neural representations. Loop soundness is separated from conditional finite-grid termination under idealized feasibility assumptions. We analyze why scaling and fixed-margin training can leave valid metrics uncertified. The scaling construction requires a scale-closed family, which the experimental fixed-floor class generally lacks. Verified slack improves the contraction rate and quantifies model-error tolerances using certified residual and Jacobian bounds. Finite-distance conclusions require connecting paths whose flows stay in the verified region. Experiments include controlled ablations, three-dimensional verification, and transfer from a fitted surrogate with known-truth residual certification. A physical swing-model study compares neural learning objectives with shared verification and invariant energy regions. Verified constant metrics certify the polynomial benchmark and smallest swing-model domain; neither result requires neural state dependence. Exploratory refinement certifies additional frozen candidates at a larger verification budget. In matched polynomial operating-grid experiments, trace normalization and margin coupling did not demonstrate a certification-count advantage within the tested budgets. These choices address specific failure mechanisms without guaranteeing optimizer convergence or general empirical superiority.
\end{abstract}


\section{Introduction}

Learning a contraction metric and proving that it works are different tasks. A small training loss on sampled states does not establish the contraction inequality throughout a domain. Conversely, a verifier with a conservative margin can reject a metric that already satisfies that inequality. Certificate-learning methods must therefore account for both the gap between sampled training and domain-wide validity and the gap between validity and numerical certifiability \cite{dawson2023}.

We study a counterexample-guided workflow for continuous-time autonomous systems that connects metric learning, sound numerical verification, and quantitative dynamical guarantees. A polynomial or neural parametrization proposes a metric; a verifier can certify its matrix inequality or return diagnostic points for retraining. Computable bounds and outward-rounded checks establish the meaning of a positive answer. Verified positive slack yields explicit contraction-rate improvements and, given global residual bounds, sufficient model-error tolerances. The implementation records the parameters, bounds, outcomes, and build provenance needed to audit successful certificates. The analysis explains how learning and certification can fail for different reasons, while the experiments delimit the evidence for the proposed training choices.

\subsection{Related work}

\noindent\textit{Contraction certificates.} Contraction analysis \cite{lohmiller1998,forni2014,tsukamoto2021tutorial,bullo2026} provides differential stability guarantees and, with suitable flow geometry, incremental trajectory bounds. Simpson-Porco and Bullo \cite{simpsonporco2014} develop an intrinsic Riemannian formulation and clarify its geometric hypotheses. Contraction under general norms also yields robustness results \cite{davydov2022noneuclidean}. Our condition remains Riemannian; the infinity norm in Remark~\ref{rem:coordinate} is used only for grid coverage. Polynomial contraction metrics can be synthesized by sum-of-squares (SOS) programming \cite{aylward2008}; rigorous numerical constructions also use meshfree collocation \cite{giesl2019,giesl2021}. Stability certificates with statistical guarantees have been learned from trajectories \cite{boffi2020}. Sampling-based verification extends inequalities from finitely many points by continuity bounds \cite{bobiti2018}. Our grid oracle follows this established principle for a matrix certificate, with computed global constants and outward-rounded inequalities.

\noindent\textit{Counterexample-guided certificate synthesis.} Counterexample-guided inductive synthesis (CEGIS) \cite{solar2006} alternates candidate generation and verification. Neural Lyapunov and controller synthesis use this architecture \cite{chang2019,abate2021}, and tools such as Fossil 2.0 implement certificate-learning loops \cite{edwards2024}. Related verification methods also certify deep reinforcement learning controllers \cite{mandal2024}. Polynomial control Lyapunov functions can also be learned from counterexamples and demonstrations, with convergence results under different learner and verifier assumptions \cite{ravanbakhsh2019}. For discrete-time systems, Li et al.\ \cite{li2025} train neural contraction metrics with projected-gradient counterexamples and perform final verification with $\alpha,\beta$-CROWN. Their condition compares local distances rather than differentiating a continuous-time matrix inequality.

\noindent\textit{Verified neural contraction metrics.} Neural representations of continuous-time contraction metrics exist \cite{tsukamoto2021ncm,sun2021}. Verification includes Lipschitz margins over grids \cite{sun2021} and post-training SMT verification \cite{fitzsimmons2024}. Fitzsimmons and Liu \cite{fitzsimmons2024} compute metrics for autonomous systems and formally verify them after training. Section~\ref{sec:application} compares an adaptation of their spectral partial-differential-inequality (PDI) loss with our learner on their physical swing model. It matches the neural metric class, training targets, initialization, computational caps, and final verifier. This controlled comparison does not reproduce their original architecture, learned invariant domain, training budget, or SMT workflow.

\noindent\textit{Verification during neural learning.} Davydov \cite{davydov2025partitioning} integrates interval bound propagation and partitioned Metzler-majorant matrix checks into joint neural controller-and-metric training, with a loss on partition-wise certificate bounds. Harapanahalli, Coogan, and Davydov \cite{harapanahalli2026} jointly train a neural controller and metric using differentiable linear bound propagation and interval matrix checks. Our learner instead trains sampled hinge losses, augments its dataset with diagnostic points, and then verifies the candidate. Verification during training and domain partitioning are established antecedents; autonomous rather than controlled synthesis is a scope distinction. Our focus is failure-mechanism analysis and computable conversion of verified slack into rate and model-error guarantees. The present experiments do not establish a performance advantage over these methods.

\noindent\textit{Learning contracting models and certificate transfer.} Jaffe et al.\ \cite{jaffe2024} build contraction structure into learned dynamics through a neural vector-field parametrization and learned coordinate changes. Here, the given or fitted vector field is held fixed during metric synthesis, and its contraction condition is certified on a prescribed compact domain. Contraction-based planning with learned dynamics has been studied with probabilistic model-error bounds and tracking tubes \cite{chou2021}. Neural Lyapunov control for unknown dynamics also transfers learned-model certificates to the unknown system under stated error assumptions \cite{zhou2022}. Our residual result concerns deterministic preservation of an autonomous contraction inequality under uniform bounds on both the residual and its Jacobian. It complements tracking-tube results; it does not infer either bound from finite data alone.

\subsection{Contributions and scope}

The principal contribution is an integrated analysis of learning and certification failures, supported by computable outward-rounded certificates and reproducible diagnostics. Sampling-and-continuity verification and CEGIS supply established building blocks \cite{bobiti2018,solar2006}; the verified-margin rate and residual estimates below are derived consequences of the contraction inequality. We organize the development into four parts. First, we give computable polynomial constants and outward-rounded grid checks, establish loop soundness, and state conditional finite-grid termination for an idealized feasible learner (Section~\ref{sec:framework}). Second, we convert verified uniform slack into an improved differential contraction rate and an explicit tolerance to model residuals (Section~\ref{sec:results}). Third, scaling and stationary-loss observations identify failure mechanisms and motivate trace normalization, margin coupling, and bounded-parameter variants. The scale-chase construction requires a scale-closed family; the fixed-floor class used in the main experiments generally does not satisfy that hypothesis. Fourth, we report controlled ablations and studies of polynomial, trigonometric, and neural representations, together with certified constants and verification costs (Section~\ref{sec:experiments}). The physical swing-model application adds a matched PDI-loss comparison and a fixed-data hinge control, with invariant energy subsets that support trajectory guarantees. Analytical constant metrics certify the polynomial benchmark and the smallest swing-model box; neural state dependence is unnecessary for those results. The matched polynomial operating-grid experiments show no certification-count advantage from normalization or margin coupling within the tested budgets. Finite trajectory-distance conclusions throughout use Remark~\ref{rem:finite}.

Here, \emph{neurosymbolic} refers to the neural-metric extension: a trainable network proposes a metric factor, while verification uses an explicit mathematical contraction condition. The symbolic role is the representation and derivation of that condition, including analytic derivatives and exact polynomial coefficient algebra where applicable. Outward-rounded interval bounds certify the resulting inequalities. The main controlled ablations use a factor linear in fixed polynomial features and study the shared counterexample-guided learner--verifier loop. Section~\ref{sec:neural} evaluates the neural parametrization separately.

The polynomial benchmark has known state-dependent and constant certificates and isolates the loop's behavior; it does not establish an advantage over SOS on polynomial systems. The pendulum study works directly with trigonometric terms. SOS can also handle the pendulum through an exact sine/cosine lift with the circle constraint; that reformulation changes coordinates and the certificate representation \cite{tedrake2026}. We restrict attention to autonomous dynamics. Controller synthesis through dual control contraction metrics $W=M^{-1}$ with their associated structural conditions \cite{manchester2017}, or through direct joint controller-and-primal-metric learning \cite{davydov2025partitioning,harapanahalli2026}, is outside the present scope.

\section{Preliminaries}
\label{sec:prelim}

Consider the system
\begin{equation}
\dot{x} = f(x), \qquad x \in \X \subset \R^n,
\label{eq:sys}
\end{equation}
with $f\in C^2$ on a neighborhood of the compact set $\X$. Write $A(x) = \partial f/\partial x\,(x)$. This assumption makes $A$ locally Lipschitz. The grid oracle requires a Lipschitz bound on its test function; Lemma~\ref{lem:certconst} supplies one for smooth polynomial parametrizations. Unless specified otherwise, vector norms are Euclidean, matrix norms $\|\cdot\|_2$ are spectral, and $\|\cdot\|_F$ is the Frobenius norm. A continuously differentiable $M : \X \to \Sym^n_{++}$ with
\begin{equation}
a I \preceq M(x) \preceq b I \quad \text{on } \X, \qquad 0 < a \le b,
\label{eq:bounds}
\end{equation}
is a contraction metric with rate $\lambda > 0$ on $\X$ if
\begin{equation}
C(x) \;:=\; \dot{M}(x) + A(x)^\top M(x) + M(x) A(x) + 2\lambda M(x) \;\preceq\; 0
\qquad \text{for all } x \in \X,
\label{eq:contraction}
\end{equation}
where $\dot{M}(x) = \sum_{i=1}^{n} \frac{\partial M}{\partial x_i}(x) f_i(x)$. A variational displacement $\delta x$ along a trajectory $x(t)$ solves $\dot{\delta x}=A(x(t))\delta x$. If \eqref{eq:contraction} holds, then $\|\delta x(t)\| \le \sqrt{b/a}\, e^{-\lambda t} \|\delta x(0)\|$ throughout any time interval during which $x(t)$ remains in $\X$ \cite{lohmiller1998}.

\begin{remark}[Finite trajectory distances]
\label{rem:finite}
Let $x_0(t)$ and $x_1(t)$ be two solutions. Suppose a piecewise continuously differentiable curve $c_0:[0,1]\to\X$ connects their initial states and its flowing family $c(t,s)$, defined by $\partial_t c=f(c)$ and $c(0,s)=c_0(s)$, remains in $\X$ over the time interval considered. Integrating the variational bound over $s$ gives
\[
\|x_1(t)-x_0(t)\| \le \sqrt{b/a}\,e^{-\lambda t}\,\operatorname{length}_{E}(c_0),
\]
where $\operatorname{length}_{E}$ is Euclidean arc length. If the straight segment is admissible, this is the displayed exponential bound with $\|x_1(0)-x_0(0)\|$ on the right. A convex forward-invariant domain is a sufficient condition. Merely keeping the two endpoint trajectories inside a compact box does not ensure the connecting family remains there. Every finite trajectory-distance conclusion below uses this condition for the relevant dynamics.
\end{remark}

The candidate metric is a parametric family $M_\theta$, $\theta \in \Theta \subseteq \R^p$, with symmetry and the bounds \eqref{eq:bounds} enforced by construction, for instance the Cholesky-type parametrization
\begin{equation}
M_\theta(x) = \alpha I + L_\theta(x)^\top L_\theta(x), \qquad \alpha > 0,
\label{eq:class}
\end{equation}
with $L_\theta(x)$ lower triangular. For a feature vector $\phi(x)\in\R^q$, each of its $n(n+1)/2$ free entries is $\ell_{ij}(x)=\theta_{ij}^{\top}\phi(x)$, giving $p=q n(n+1)/2$ parameters. The planar polynomial studies use $\phi=(1,x_1,x_2)$ and therefore nine parameters. The lower bound $a = \alpha$ holds for every $\theta$; the upper bound $b$ follows from bounding $\|L_\theta\|$ on the compact $\X$. We write $C_\theta(x)$ for the matrix in \eqref{eq:contraction} evaluated at $M_\theta$, and
\begin{equation}
g_\theta(x) := \lmax(C_\theta(x)),
\qquad
\tilde g_\theta(x) := \frac{\lmax(C_\theta(x))}{\tr M_\theta(x)}.
\label{eq:tests}
\end{equation}
Both functions are sign-equivalent to \eqref{eq:contraction} pointwise since $\tr M_\theta > 0$. We call $\tilde g_\theta$ the normalized test function. Section~\ref{sec:results} explains the scale invariance and the choice of the trace denominator.

\noindent\textit{Notation.} The raw and normalized tests are $g$ and $\tilde g$; $L_h$ is a certified test Lipschitz bound and $\tau$ is a grid covering radius. We distinguish true uniform slack $s_*=-\sup_\X h$, computed lower slack $\underline s$, and raw matrix margin $\eta$. The gap $\Gamma$ is the oracle residual. The trace lower bound is $\underline v$, $\widehat G$ bounds $|\tilde g|$, and $D_k$ bounds a matrix partial derivative. The coefficient bound is $\mathrm{sb}(\cdot)$. The metric bounds are $a,b$, the metric-derivative constant is $\mu_1$, and residual bounds are $\delta_0,\delta_1$. The improved model and true-system rates are $\lambda'$ and $\lambda''$.

\section{Counterexample-guided synthesis framework}
\label{sec:framework}

\subsection{Verification oracle}

\begin{definition}[Verification problem]
Given $f$, $\X$, $\lambda$, and a candidate $\theta$, certify $g_\theta\le0$ throughout $\X$, or return diagnostic points at which certification failed. A point is a true counterexample only when a positive lower bound proves $g_\theta(x)>0$.
\end{definition}

An oracle is sound if its positive answers are correct. The grid oracle below is sound but not complete: it can reject a candidate that satisfies the contraction condition when the true slack falls below the grid margin, so Definition~\ref{def:oracle} decides only up to that margin, and we return to this incompleteness in Remark~\ref{rem:complete}. We use a grid oracle following sampling-and-continuity verification \cite{bobiti2018,sun2021}, stated here for an arbitrary test function $h \in \{g_\theta, \tilde g_\theta\}$.

\begin{definition}[Grid oracle]
\label{def:oracle}
Let $h : \X \to \R$ be Lipschitz in the Euclidean norm with constant $L_h$, and let $\X_\tau = \{x_1, \dots, x_N\} \subset \X$ satisfy $\sup_{x \in \X} \min_i \|x - x_i\| \le \tau$. The oracle certifies if $h(x_i) \le -L_h \tau$ for all $i$, and otherwise returns the grid points violating this margin.
\end{definition}

\begin{proposition}[Soundness]
\label{prop:grid}
If the grid oracle certifies with test function $g_\theta$ or $\tilde g_\theta$, then \eqref{eq:contraction} holds on $\X$ and the system \eqref{eq:sys} is contracting on $\X$ with rate $\lambda$ in the metric $M_\theta$.
\end{proposition}

\begin{proof}
The map $S \mapsto \lmax(S)$ is $1$-Lipschitz on symmetric matrices in the spectral norm by Weyl's inequality, so $g_\theta$ is Lipschitz whenever $x \mapsto C_\theta(x)$ is, and $\tilde g_\theta$ is Lipschitz as a quotient of Lipschitz functions with denominator bounded below by $n\alpha$ on the compact $\X$; Lemma~\ref{lem:certconst} furnishes an explicit constant $L_h$ for the polynomial class; Section~\ref{sec:neural} treats the neural case. Fix $x \in \X$ and a grid point $x_i$ with $\|x - x_i\| \le \tau$. Then $h(x) \le h(x_i) + L_h \tau \le 0$. Since $h$ is sign-equivalent to \eqref{eq:contraction} pointwise, the condition holds on $\X$ and gives the variational conclusion of Section~\ref{sec:prelim}. Finite trajectory-distance bounds use Remark~\ref{rem:finite}.
\end{proof}

\begin{corollary}[Uniform margin from grid margin]
\label{cor:margin}
If additionally $h(x_i) \le -L_h\tau - \eta$ for all $i$ with $\eta > 0$, then $h(x) \le -\eta$ for all $x \in \X$.
\end{corollary}

\begin{proof}
Identical, carrying the extra $-\eta$ through the Lipschitz estimate.
\end{proof}

\subsection{Certified constants for the polynomial class}
\label{sec:certconst}

The oracle of Definition~\ref{def:oracle} consumes a Lipschitz constant $L_h$, and its soundness rests on that constant being valid. For the class \eqref{eq:class} with polynomial features and polynomial $f$, every entry of $C_\theta$ and of $\partial C_\theta/\partial x_k$ is a polynomial in $x$ whose coefficients are explicit in $\theta$, so $L_h$ is computed, not estimated. We record the bounds the implementation uses.

For a polynomial $p$, write $\mathrm{sb}(p)$ for a certified absolute coefficient bound after a fixed subbox subdivision: recenter $p$ on each subbox, sum the absolute monomial contributions over its half-width, and take the largest bound. Lemma~\ref{lem:coeff} in Appendix~\ref{app:coeff} proves this bound and its refinement property.

\begin{lemma}[Certified Lipschitz and class constants]
\label{lem:certconst}
Let $\X=[-B,B]^n$, let $f$ and the features of \eqref{eq:class} be polynomial, and write $c_{ij}$ and $m_{ij}$ for the entries of $C_\theta$ and $M_\theta$. Define
\[
D_k=\left(\sum_{i,j=1}^n\mathrm{sb}(\partial_k c_{ij})^2\right)^{1/2},
\qquad
v=\tr M_\theta,
\qquad
V_k=\mathrm{sb}(\partial_k v).
\]
Then $\sup_\X\|\partial_k C_\theta\|_F\le D_k$, and the following bounds hold.
(i) The raw test is Euclidean Lipschitz with
\[
L_g\le\left(\sum_{k=1}^nD_k^2\right)^{1/2}.
\]
(ii) Choose certified bounds
\[
0<n\alpha\le\underline v\le\inf_\X v,
\qquad
\widehat G\ge\sup_\X|\tilde g_\theta|.
\]
The lower bound $\underline v$ can be $n\alpha$ or a tighter coefficient lower bound. One valid $\widehat G$ is the largest, over subboxes, of $\sqrt{\mathrm{sb}(\|C_\theta\|_F^2)}$ divided by a certified positive local lower bound on $v$. The trace and normalized tests satisfy
\[
L_v\le\left(\sum_kV_k^2\right)^{1/2},
\qquad
L_{\tilde g}\le\frac{L_g+\widehat G L_v}{\underline v}.
\]
(iii) The class bounds \eqref{eq:bounds} hold with $a=\alpha$ and
\[
b\le\left(\sum_{i,j=1}^n\mathrm{sb}(m_{ij})^2\right)^{1/2}.
\]
(iv) The metric-derivative constant of Theorem~\ref{thm:residual} satisfies
\[
\mu_1\le\left(\sum_{k=1}^n\sum_{i,j=1}^n\mathrm{sb}(\partial_k m_{ij})^2\right)^{1/2}.
\]
\end{lemma}

The entrywise Frobenius sums count each off-diagonal entry twice, recovering the planar implementation. The formulas hold for arbitrary $n$; the original polynomial and pendulum grid implementations instantiate $n=2$. Appendix~\ref{app:coeff} gives the proof.

\begin{remark}[Coordinatewise covering bound]
\label{rem:coordinate}
The proof gives $|g(x)-g(y)|\le\sum_kD_k|x_k-y_k|$. For the normalized test replace $D_k$ by $E_k=(D_k+\widehat G V_k)/\underline v$. On an axis grid with maximum cell half-width $r_k$ in coordinate $k$, the required margin can therefore be $\sum_kD_kr_k$ or $\sum_kE_kr_k$. For equal half-width $r$, this is the infinity-norm covering bound and is no larger than the corresponding Euclidean bound $r\sqrt n\,(\sum_kD_k^2)^{1/2}$, or its normalized counterpart. The reported baseline experiments retain their Euclidean oracle; changing this bound defines a different verification regime. The infinity norm here describes coverage of the state box, not a change to the Riemannian contraction condition.
\end{remark}

\begin{lemma}[Grid sufficiency]
\label{lem:gridsuff}
Let $h$ be Lipschitz with constant $L_h \ge 0$ on $\X$ and let $\X_\tau$ have covering radius $\tau$. Write $s_* := -\sup_\X h$ for the true uniform slack and $\underline s := -\max_i h(x_i)-L_h\tau$ for its grid-based lower bound. If $s_* \ge L_h\tau+\eta$ for some $\eta\ge0$, then every grid point satisfies $h(x_i)\le-(L_h\tau+\eta)$ and the oracle of Definition~\ref{def:oracle} certifies. Whenever the oracle certifies, $0\le\underline s\le s_*$, so $h\le-\underline s$ on $\X$. For $\X=[-B,B]^n$ and a uniform axis grid with $N_{\mathrm{ax}}\ge2$ points per axis, the covering radius is $\tau=B\sqrt n/(N_{\mathrm{ax}}-1)$. If $s_*>0$, a sufficient axis count is
\[
N_{\mathrm{ax}} \;\ge\; 1 + \frac{B\sqrt{n}\, L_h}{s_*},
\]
with $N=N_{\mathrm{ax}}^n$ points in total. If $L_h=0$, any nonempty grid certifies exactly when $s_*\ge0$. If $s_*=0<L_h$, this sufficiency argument gives no finite axis count, a quantitative instance of the oracle incompleteness in Remark~\ref{rem:complete}.
\end{lemma}

\begin{proof}
Every grid value is at most $\sup_\X h=-s_*$, which proves the first implication. For each $x\in\X$, choose a grid point within $\tau$. The Lipschitz estimate gives $h(x)\le\max_i h(x_i)+L_h\tau=-\underline s$, hence $s_*\ge\underline s$. Certification is exactly $\underline s\ge0$. The uniform-grid covering radius is half its cell diagonal; solving $L_h\tau\le s_*$ for $N_{\mathrm{ax}}$ when $s_*>0$ gives the stated sufficient count. When $L_h=0$, the Lipschitz inequality makes $h$ constant on $\X$, which proves the remaining case.
\end{proof}

\noindent\textit{Numerical guarantees.} Stored binary64 parameters denote exact real inputs. Polynomial coefficient construction is exact rational arithmetic; conversion, elementary bounds, and transcendental evaluations are outward-rounded. A positive answer requires directed-rounded inequalities over every relevant grid point, and reported certified slack is rounded downward. Rejection can be diagnostic or inconclusive and does not prove a contraction violation. Appendix~\ref{app:numerics} records the arithmetic assumptions, interpolation bounds, and fail-closed policy.

\subsection{The CEGIS-C loop}

Algorithm~\ref{alg:cegis} states the loop in its final form: the learner minimizes a hinge loss on the normalized test function with a margin coupled to the oracle's requirement. Sections~\ref{sec:results} and \ref{sec:experiments} explain how these choices address scaling and fixed-margin failure mechanisms. Remark~\ref{rem:complete} states the limitations on convergence.

\begin{algorithm}[t]
\caption{CEGIS-C (normalized, margin-coupled)}
\label{alg:cegis}
\begin{algorithmic}[1]
\Require dynamics $f$, domain $\X$, rate $\lambda$, family $\{M_\theta\}$, initial parameter $\theta_0$, grid $\X_\tau$, base margin $\varepsilon_0 > 0$, coupling factor $\gamma > 1$, maximum returned-point count $q_c$, perturbation count $m$, Gaussian standard deviation $\sigma$, iteration cap $K$
\State $\theta \gets \theta_0$; \quad $\D \gets$ initial samples in $\X$; \quad $\varepsilon \gets \varepsilon_0$
\For{$k = 0, 1, \dots, K$}
  \State $(\textsf{status}, \mathcal{C}, L_h, \underline s) \gets \mathsf{Ver}(\tilde g_\theta, \X_\tau)$ \Comment{$\underline s = -\max_i \tilde g_\theta(x_i) - L_h\tau$}
  \If{$\textsf{status} = \textsf{certified}$} \Return $(\theta, \textsf{certified}, L_h, \underline s, a, b)$ \Comment{$a = \alpha$; certified metric bound $b$}
  \EndIf
  \If{$k = K$} \Return $(\theta, \textsf{failure})$ \Comment{at most $K$ retraining rounds}
  \EndIf
  \State $\varepsilon \gets \max\{\varepsilon_0,\; \gamma L_h \tau\}$ \Comment{margin coupling}
  \State $\mathcal C \gets$ up to $q_c$ worst margin violators
  \State $\D \gets \D \cup \mathcal{C} \cup \mathsf{Perturb}(\mathcal{C}, m,\sigma)$ \Comment{clip perturbations to $\X$}
  \State $\theta \gets \textsc{Learn}(\theta, \D, \varepsilon)$ \Comment{fixed-budget subgradient steps on $\frac{1}{|\D|}\sum_{x \in \D} \max\{0,\, \tilde g_\theta(x) + \varepsilon\}$, warm-started at $\theta$}
\EndFor
\State \Return $(\theta, \textsf{failure})$
\end{algorithmic}
\end{algorithm}

A successful run returns the certified metric together with the constants a downstream guarantee consumes: the oracle constant $L_h$, the certified lower bound $\underline s$ on uniform normalized slack from Lemma~\ref{lem:gridsuff}, and the class bounds $a=\alpha$ and $b$. The polynomial implementation obtains $b$ from Lemma~\ref{lem:certconst}(iii); the neural implementation uses outward interval bounds on the metric entries over the whole domain. Theorems~\ref{thm:rate} and \ref{thm:residual} read the raw uniform margin $\eta$ from $\underline s$ through $\tr M_\theta\ge na$, as Remark~\ref{rem:eta} makes precise.

The learner $\textsc{Learn}$ is a fixed-budget subgradient method, Adam in our experiments. Parameters are warm-started at the current $\theta$, while Adam first- and second-moment estimates are reset to zero at each retraining round. We write it as a finite-step update rather than an exact minimizer because the stall analysis of Proposition~\ref{lem:stall} concerns gradient steps, and an exact argmin is non-unique under the stall, where every $\theta$ of zero loss qualifies.

\begin{proposition}[Soundness of CEGIS-C]
\label{prop:cegis}
If the oracle is sound and Algorithm~\ref{alg:cegis} returns $(\theta, \textsf{certified})$, then $M_\theta$ is a contraction metric with rate $\lambda$ on $\X$. Every variational displacement along a trajectory of \eqref{eq:sys} remaining in $\X$ satisfies $\|\delta x(t)\| \le \sqrt{b/a}\, e^{-\lambda t}\|\delta x(0)\|$. Finite trajectory-distance bounds require the connecting-family condition of Remark~\ref{rem:finite}.
\end{proposition}

\begin{proof}
Immediate from Proposition~\ref{prop:grid}: the loop changes which candidate reaches the oracle, while a positive answer has the same meaning.
\end{proof}

\begin{remark}[Soundness without completeness]
\label{rem:complete}
Proposition~\ref{prop:cegis} establishes soundness, not completeness or termination of the implemented finite-budget learner. The grid oracle can reject a metric with positive true slack below $L_h\tau$. Lemma~\ref{lem:gridsuff} guarantees certification when true slack clears the grid margin. The conditional result below applies to a different, idealized learner; it does not establish relative completeness of Algorithm~\ref{alg:cegis}.
\end{remark}

\begin{proposition}[Conditional finite-grid termination]
\label{prop:termination}
Fix a grid $Q=\{x_1,\ldots,x_N\}$ with radius $\tau$, a family $\Theta$, and valid exact-arithmetic bounds $L_h(\theta)\le\overline L$ for every $\theta\in\Theta$. Fix a training margin $\varepsilon\ge\overline L\tau$. Assume that (i) one $\theta_*\in\Theta$ satisfies $h_{\theta_*}(x)\le-\varepsilon$ for every point that may enter the dataset; (ii) before the initial oracle call and after each augmentation, a feasible learner returns $\theta\in\Theta$ satisfying $h_\theta(x)\le-\varepsilon$ at every accumulated point; and (iii) a rejecting oracle returns at least one grid point satisfying the genuine strict margin violation $h_\theta(x_i)>-L_h(\theta)\tau$. Then the loop certifies after at most $N-|\D_0\cap Q|$ augmentations. If all data lie in $Q$, realizability need only hold on $Q$. With off-grid initial samples or perturbations, realizability on all of $\X$ is sufficient.
\end{proposition}

\begin{proof}
Realizability makes every accumulated feasibility problem nonempty. At each oracle call, any $x_i\in\D\cap Q$ satisfies $h_\theta(x_i)\le-\varepsilon\le-\overline L\tau\le-L_h(\theta)\tau$. Every returned genuine margin violator is therefore a new grid point. Each rejection strictly increases $|\D\cap Q|$. Once the grid is exhausted, every grid inequality holds and the oracle certifies. If the initial parameter is not obtained from the feasible learner, one additional initial augmentation may be needed.
\end{proof}

For a fixed positive floor and a bounded parameter set $\|\theta\|\le R$, Lemma~\ref{lem:certconst} supplies a uniform finite bound: bound its coefficient polynomials jointly on $\X$ and $|\theta_j|\le R$, retain $v\ge n\alpha$, and substitute these bounds into $L_{\tilde g}$. This gives a computable $\overline L(R)$, but a bounded set can still be unrealizable at its required margin. Projection after unconstrained training need not preserve zero loss. Bounded Adam experiments therefore test a design choice, not the feasible-learner hypothesis of Proposition~\ref{prop:termination}.

The proposition uses an exact oracle. A directed-rounded implementation can return an inconclusive point already in the dataset, so fresh-point progress does not follow from soundness alone. Extending the conclusion requires an additional uniform numerical enclosure-error bound and a training buffer that exceeds it, or exact resolution of these comparisons. Similarly, a coupled-margin variant needs verified control $L_h(\theta_{k+1})\le\gamma L_h(\theta_k)$ plus feasibility of every new margin; a parameter trust region alone does not ensure this bound. The finite-grid bound can be enormous and says nothing about practical training time.

When the certified grid slack is strict, Corollary~\ref{cor:margin} and Remark~\ref{rem:eta} provide the improved rate of Theorem~\ref{thm:rate}. The next section separates failure to find a valid metric from failure to certify an already valid one.

\section{Main results}
\label{sec:results}

\subsection{Scale invariance and normalization}

The contraction condition \eqref{eq:contraction} is homogeneous of degree one in the metric. This observation separates changes in scale from changes in certifiable shape.

\begin{assumption}[Scale-equivariant oracle bound]
\label{ass:scale}
On the raw test, the exact bounding rule obeys $L_g(cM)=cL_g(M)$ for $c>0$. On the normalized test it obeys $L_{\tilde g}(cM)=L_{\tilde g}(M)$. These are properties of the bounding rule, not of an arbitrary choice of valid upper bounds. For the normalized coefficient bound, all auxiliary trace and test bounds must be chosen consistently under scaling. Finite-precision enclosures need not preserve the equalities bit for bit.
\end{assumption}

\begin{lemma}[Scale invariance of certifiability]
\label{lem:scale}
Fix $x \mapsto M(x)$ and $c > 0$, and let $C(\cdot\,; M)$ denote the matrix in \eqref{eq:contraction}. Then $C(x; cM) = c\, C(x; M)$ for all $x$, hence $\lmax(C(x; cM)) = c\,\lmax(C(x; M))$, and any Lipschitz constant $L$ of $x \mapsto \lmax(C(x;M))$ scales to $cL$. Consequently, under the raw-test part of Assumption~\ref{ass:scale}, which the exact coefficient construction satisfies when applied to the scaled matrix itself, for every grid $\X_\tau$ the metric $cM$ passes the grid oracle on the test function $g$ if and only if $M$ does: certifiability is a property of the metric's shape, invariant under positive scaling. The normalized test function $\tilde g$ of \eqref{eq:tests} is itself invariant: $\tilde g(x; cM) = \tilde g(x; M)$.
\end{lemma}

\begin{proof}
Each term of \eqref{eq:contraction} is linear in $M$: $\dot{(cM)} = c\dot M$, $A^\top (cM) + (cM) A = c(A^\top M + MA)$, and $2\lambda (cM) = 2c\lambda M$. Eigenvalues are positively homogeneous, giving the second claim, and $|c\,g(x) - c\,g(y)| = c\,|g(x) - g(y)|$ gives the third, with the converse scaling $1/c$ showing the constant is exactly $cL$ when $L$ is tight. Under the scale-equivariance hypothesis the oracle's certificate reads $c\,g(x_i) \le -cL\tau$, which is the original inequality divided by $c$. For the last claim, $\tr(cM) = c \tr M$ cancels the numerator's scaling.
\end{proof}

\begin{definition}[Certifiability gap]
\label{def:gap}
For a candidate metric $M$ and grid $\X_\tau$ with oracle constant $L_h$, the certifiability gap on the test $h \in \{g, \tilde g\}$ is
\[
\Gamma(M) \;:=\; \max_i h(x_i) + L_h\tau,
\]
the worst grid value of $h$ minus the oracle's required margin $-L_h\tau$; the oracle certifies exactly when $\Gamma(M) \le 0$. On the raw test the gap is scale-covariant, $\Gamma(cM) = c\,\Gamma(M)$ for $c > 0$ by Lemma~\ref{lem:scale}, while on the normalized test it is scale-invariant provided the oracle bound also respects that invariance.
\end{definition}

\begin{corollary}[The scale chase]
\label{cor:chase}
Fix a grid $\X_\tau$ and an initial metric $M_0$ whose raw test satisfies $g_0(x)<0$ throughout the compact domain $\X$. Thus $M_0$ is already a valid contraction metric at the target rate. Suppose the raw oracle bound is scale-equivariant with $L_0>0$ and the initial gap is $\Gamma(M_0)>0$. Assume the family is closed under upward scaling: every $cM_0$ with $c\ge1$ belongs to the family. This hypothesis generally excludes the fixed-floor implementation \eqref{eq:class}. At a rejected candidate $M_k$, let the next training requirement be $g(x)\le-\varepsilon_k$ on the augmented data, where $\varepsilon_k=\gamma L_k\tau$ and $\gamma>1$. Then there exists an infinite sequence of learner responses obtained by scaling $M_0$ alone. Every response satisfies its training requirement, every candidate remains uncertified, and $L_k\to\infty$.
\end{corollary}

\begin{proof}
Continuity and compactness give $m_0:=-\sup_\X g_0>0$. Write $M_k=c_kM_0$ with $c_0=1$. After the oracle returns its candidates and the data are augmented, choose $c_{k+1}\ge\max\{2c_k,\varepsilon_k/m_0\}$. Then $g_{k+1}(x)=c_{k+1}g_0(x)\le-\varepsilon_k$ on all of $\X$, hence on every future training point as well. Lemma~\ref{lem:scale} gives $L_k=c_kL_0$ and $\Gamma(M_k)=c_k\Gamma(M_0)>0$. Thus the oracle keeps rejecting while $c_k\ge2^k$ and $L_k\to\infty$. This constructs admissible responses; it does not require a particular optimizer to choose them.
\end{proof}

Negativity on the current dataset alone suffices for a single fixed-data scaling step, but cannot support the corollary after new points are added: scaling preserves the sign of every raw test value. The uniform negativity hypothesis supplies the missing control over future data. The positive bound and initial positive gap ensure a growing, noncertifying sequence.

The implemented class \eqref{eq:class} with fixed $\alpha$ is generally not closed under scaling, since $cM_\theta=c\alpha I+cL_\theta^\top L_\theta$ need not admit a factorization $\alpha I+L'^\top L'$ with $L'$ linear in the same features. Corollary~\ref{cor:chase} therefore does not by itself establish a scale chase for this fixed-floor class. If $L_\theta^\top L_\theta$ is uniformly positive definite, scaling that term makes the fixed floor relatively negligible; establishing an infinite learner sequence for this limit would require additional hypotheses. Section~\ref{sec:ablation} separates the fixed-floor ablation from a constructive scale-closed illustration.

The corollary identifies an admissible failure sequence: a gradient-based learner is not forced to scale, and on our benchmark it reaches a certifiable shape rather than inflating pure scale. The coupled-margin divergence that Section~\ref{sec:experiments} does exhibit is driven by the metric's shape through its Lipschitz constant, a different mechanism from pure scale inflation. Lemma~\ref{lem:scale} motivates a remedy for the scaling degeneracy. Both this construction and the stall below involve already valid metrics that the oracle cannot certify at its current margin. Tightening the bound or refining the grid can therefore help without changing the metric. Training for a better slack-to-Lipschitz ratio seeks easier certification, rather than merely feasibility of the contraction inequality. The normalized test function $\tilde g$ is scale-invariant, so scaling alone cannot lower its training values. If the oracle bound is also invariant, reducing the certifiability gap requires changing the metric's shape. Scale invariance removes this degeneracy; the lemma does not require the trace quotient in particular or rule out other ways to control metric scale. The generalized-eigenvalue normalization $\lmax(M^{-1/2} C M^{-1/2})$ is equally scale-invariant and reads directly as the pointwise rate deficit. We use the trace normalization because its denominator $\tr M_\theta$ is a low-degree polynomial with a certified Lipschitz constant and the exact lower bound $n\alpha$, which keeps the oracle's constant computable by Lemma~\ref{lem:certconst}, whereas the generalized-eigenvalue quotient would entangle the denominator's Lipschitz analysis with a matrix inverse. Algorithm~\ref{alg:cegis} therefore trains on $\tilde g$, with the trace chosen for tractable Lipschitz bounds.

\subsection{The stall mode of fixed-margin learners}

\begin{proposition}[Fixed-margin stationary updates]
\label{lem:stall}
Fix a test function $h_\theta \in \{g_\theta, \tilde g_\theta\}$ that is locally Lipschitz in $\theta$, a grid $\X_\tau$ with oracle constant $L_h$, and a fixed learner margin $0<\varepsilon < L_h \tau$. Suppose the current candidate satisfies
\begin{equation}
\sup_{x \in \X} h_\theta(x) \le -\varepsilon
\qquad \text{and} \qquad
\max_i h_\theta(x_i) > -L_h \tau .
\label{eq:stallband}
\end{equation}
Then the candidate is uncertified and the hinge loss $\max\{0,h_\theta+\varepsilon\}$ vanishes identically on $\X$. For every finite dataset $\D\subset\X$, strict inequality in the first condition makes this loss locally constant at $\theta$. At the boundary, zero belongs to its Clarke subdifferential. A learner that selects this zero subgradient and has zero optimizer state can keep $\theta$ fixed after every data augmentation. Thus the loop can remain at this non-certified point.
\end{proposition}

\begin{proof}
The second inequality in \eqref{eq:stallband} is the negation of the oracle's certificate. The first makes the empirical hinge loss zero on every finite dataset in $\X$, including all points the oracle can ever return. For strict inequality, continuity in $\theta$ and finiteness of $\D$ give a neighborhood with zero loss. At the boundary, the nonnegative locally Lipschitz loss attains a global minimum, so its Clarke subdifferential contains zero. Selecting zero at each update with zero optimizer state leaves $\theta$ unchanged, and \eqref{eq:stallband} persists after augmentation. A different boundary subgradient selection or nonzero momentum can move the parameter; the conclusion asserts the existence of a stalled update sequence.
\end{proof}

The interval $(-L_h\tau, -\varepsilon]$ in \eqref{eq:stallband} is the stall band. It is nonempty exactly when $\varepsilon < L_h\tau$, that is, when the training requirement is weaker than the verification margin. The coupling step of Algorithm~\ref{alg:cegis} sets the next training margin above the current oracle requirement. This removes the fixed-margin stall band for that candidate when $L_h\tau > 0$. Retraining changes $L_h$, so the step does not guarantee certification. Lemma~\ref{lem:scale} motivates combining this step with normalization to remove pure scale inflation as a way to satisfy the training margin. These observations justify these design choices as remedies for the stated failure mechanisms, not as necessary or sufficient conditions for convergence.

\subsection{Slack converts to guarantees}

Verified uniform slack improves the differential rate and gives a sufficient model-error budget.

\begin{theorem}[Rate improvement from verified margin]
\label{thm:rate}
Suppose a sound verification method establishes $\lmax(C_\theta(x))\le-\eta$ throughout $\X$ for a uniform raw margin $\eta>0$, for example the grid oracle under Corollary~\ref{cor:margin} or the adaptive oracle of Proposition~\ref{prop:adaptive}. Then $M_\theta$ is a contraction metric for \eqref{eq:sys} on $\X$ with rate
\[
\lambda' = \lambda + \frac{\eta}{2b},
\]
and every variational displacement along a trajectory remaining in $\X$ satisfies
\[
\|\delta x(t)\| \le \sqrt{b/a}\, e^{-\lambda' t} \|\delta x(0)\|.
\]
The same rate applies to finite trajectory-distance bounds under Remark~\ref{rem:finite}.
\end{theorem}

\begin{proof}
From $M_\theta \preceq bI$ we get $I \succeq M_\theta / b$, hence $-\eta I \preceq -(\eta/b) M_\theta$. Then
\[
\dot M_\theta + A^\top M_\theta + M_\theta A + 2\lambda M_\theta \;\preceq\; -\eta I \;\preceq\; -\tfrac{\eta}{b} M_\theta,
\]
which rearranges to $\dot M_\theta + A^\top M_\theta + M_\theta A + 2\lambda' M_\theta \preceq 0$ with $\lambda' = \lambda + \eta/(2b)$. Apply \cite{lohmiller1998}.
\end{proof}

\begin{remark}[Uniform margin delivered to the theorems]
\label{rem:eta}
If the loop certifies the normalized test with lower slack $\underline s\ge0$ and a certified trace lower bound $\underline v$, then
\[
g_\theta=\tilde g_\theta\tr M_\theta\le-\underline s\,\underline v\quad\text{on }\X.
\]
For $\underline s>0$, the raw margin $\eta=\underline v\,\underline s$ and rate $\lambda'=\lambda+\underline v\,\underline s/(2b)$ follow. The universal choice $\underline v=na$ recovers the original conversion; a tighter trace bound can improve it. Numerical products and reported rates are rounded downward. Each table row identifies the conversion used. These are computed lower bounds, not the unknown true slack $s_*$. When $\underline s=0$, the base rate remains certified but the strict-margin improvement does not follow.
\end{remark}

The next theorem transfers a certificate for a surrogate vector field to the true dynamics under certified residual bounds.

\begin{theorem}[Residual-aware certificate]
\label{thm:residual}
Let the true dynamics be $\dot x = f(x) + d(x)$ on $\X$, where $f$ is the model used for synthesis, $d\in C^1$ on a neighborhood of $\X$, and the residual satisfies $\|d(x)\| \le \delta_0$ and $\|\partial d/\partial x\,(x)\|_2 \le \delta_1$ on $\X$. Let $M_\theta$ be certified for the model with uniform margin $\eta$ as in Theorem~\ref{thm:rate}, and define the metric-derivative constant
\[
\mu_1 := \sup_{x \in \X} \Big( \textstyle\sum_{k=1}^{n} \big\| \frac{\partial M_\theta}{\partial x_k}(x) \big\|_2^2 \Big)^{1/2}.
\]
If $\eta > \mu_1 \delta_0 + 2 b \delta_1$, then $M_\theta$ is a contraction metric for the true dynamics on $\X$ with rate
\[
\lambda'' = \lambda + \frac{\eta - \mu_1 \delta_0 - 2 b \delta_1}{2b},
\]
and every variational displacement along a trajectory of the true system remaining in $\X$ satisfies $\|\delta x(t)\| \le \sqrt{b/a}\, e^{-\lambda'' t} \|\delta x(0)\|$. Finite trajectory-distance bounds at this rate require Remark~\ref{rem:finite} for the true flow $f+d$.
\end{theorem}

\begin{proof}
Write $D(x) = \partial d/\partial x\,(x)$. The contraction matrix of the true dynamics in the metric $M_\theta$ is
\[
C^{\mathrm{true}}_\theta(x)
= \underbrace{\dot M_\theta\big|_f + A^\top M_\theta + M_\theta A + 2\lambda M_\theta}_{C_\theta(x)}
\;+\; \sum_{k=1}^{n} \frac{\partial M_\theta}{\partial x_k}(x)\, d_k(x)
\;+\; D^\top M_\theta + M_\theta D,
\]
since the derivative of $M_\theta$ along the true flow is $\dot M_\theta\big|_f + \sum_k (\partial M_\theta/\partial x_k) d_k$ and the true Jacobian is $A + D$. By Weyl's inequality and submultiplicativity,
\[
\lmax\!\big(C^{\mathrm{true}}_\theta(x)\big)
\le \lmax(C_\theta(x))
+ \Big\| \sum_k \frac{\partial M_\theta}{\partial x_k} d_k \Big\|_2
+ 2\,\|M_\theta\|_2 \|D\|_2 .
\]
For the middle term, the triangle inequality and Cauchy--Schwarz on the vector of norms give
\[
\Big\| \sum_k \frac{\partial M_\theta}{\partial x_k} d_k \Big\|_2
\le \sum_k \Big\|\frac{\partial M_\theta}{\partial x_k}\Big\|_2 |d_k|
\le \Big( \sum_k \Big\|\frac{\partial M_\theta}{\partial x_k}\Big\|_2^2 \Big)^{1/2} \|d\|_2
\le \mu_1 \delta_0 .
\]
Combining with $\lmax(C_\theta) \le -\eta$, $\|M_\theta\|_2 \le b$, and $\|D\|_2 \le \delta_1$ yields $\lmax(C^{\mathrm{true}}_\theta(x)) \le -(\eta - \mu_1\delta_0 - 2b\delta_1) < 0$ on $\X$. Apply Theorem~\ref{thm:rate} with $\eta$ replaced by $\eta - \mu_1\delta_0 - 2b\delta_1$.
\end{proof}

\begin{remark}
Theorem~\ref{thm:residual} gives the explicit robustness budget $\mu_1\delta_0+2b\delta_1<\eta$. If $\mu_1=0$, the residual magnitude contributes no cost to this sufficient inequality. In a pipeline where $f$ is identified from data with statistical or deterministic residual bounds, the theorem specifies how much verification slack the synthesis stage must achieve for the certificate to transfer to the true plant. The quantity $\mu_1$ is computable for the class \eqref{eq:class} since the metric derivatives are explicit in the features. The Jacobian bound $\delta_1$ is the more demanding input for a learned surrogate. A Lipschitz bound on the surrogate alone does not bound the residual Jacobian. One needs additional information about the true vector field, a direct bound on $\partial(f_{\mathrm{true}}-f)/\partial x$, or a statistical derivative-error bound with its stated confidence. Separate Jacobian bounds on the true field and a spectrally constrained surrogate yield a conservative residual bound by the triangle inequality. Setting $\delta_1 = 0$ requires a residual with zero Jacobian; a magnitude bound $\delta_0$ alone does not justify it.
\end{remark}

\begin{corollary}[Computable polynomial residual budgets]
\label{cor:polyresidual}
If $d$ is a known polynomial residual on the box, valid upper bounds in Theorem~\ref{thm:residual} are
\[
\delta_0=\sqrt{\mathrm{sb}\!\left(\sum_i d_i^2\right)},
\qquad
\delta_1=\sqrt{\mathrm{sb}\!\left(\sum_{i,j}(\partial_jd_i)^2\right)}.
\]
The second uses the Frobenius norm to upper-bound the Jacobian spectral norm. Together with Lemma~\ref{lem:certconst}, these make every term of the transfer test computable. This requires knowledge of the residual polynomial; finite fitting error alone does not provide it.
\end{corollary}

\begin{proof}
Apply Lemma~\ref{lem:coeff} to the squared vector and Frobenius norms, then use $\|\partial d/\partial x\|_2\le\|\partial d/\partial x\|_F$.
\end{proof}

\section{Computational illustrations}
\label{sec:experiments}

\subsection{Benchmark with known ground truth}
\label{sec:polynomialbenchmark}

We construct a benchmark whose certified answer is known. Let $\varphi(x) = (x_1,\, x_2 + x_1^2)$ and define $f$ by requiring $z = \varphi(x)$ to obey $\dot z = -z$. This gives the polynomial system
\begin{equation}
f(x) = \big( -x_1,\; -(x_2 - x_1^2) \big), \qquad \X = [-2, 2]^2 .
\label{eq:bench}
\end{equation}
The system contracts with rate $1$ in the pullback metric $M^*=J_\varphi^\top J_\varphi$. Its lower-triangular factor has entries linear in $x$, so $\alpha I+cM^*$ belongs to \eqref{eq:class}. To check realizability explicitly, set $m_B=(2+4B^2)^{-1}$. Since $\det M^*=1$ and $\tr M^*\le2+4B^2$ on $[-B,B]^2$, $\lambda_{\min}(M^*)\ge m_B$. For $0<\lambda<1$, Weyl's inequality gives
\[
\lmax\!\left(C_{\alpha I+cM^*}(x)\right)
\le -2c(1-\lambda)m_B+2\alpha\max\{0,\lambda-1+B\}.
\]
Thus $c>\alpha\max\{0,\lambda-1+B\}/((1-\lambda)m_B)$ suffices for strict contraction. The generated bound for the reported domain is $m_B\ge\revPullbackMinEigLower$ and the sufficient threshold is $c>\revPullbackThreshold$. These are conservative analytic bounds, not fitted values.

For $M=\beta I$, $C=\beta(A^\top+A+2\lambda I)$. The symmetric part $(A+A^\top)/2$ has eigenvalues $-1\pm|x_1|$, so $C$ has eigenvalues $2\beta(\lambda-1\pm|x_1|)$. At the target rate $\lambda=0.4$, it is indefinite exactly when $|x_1|>1-\lambda=0.6$. The class therefore contains strictly contracting shapes, while the identity-scaled initialization fails on part of the box. Certification by a particular coarse-grid margin is a stronger feasibility requirement and is not implied by this construction.

\noindent\textit{Analytical constant-metric baseline.}
A nonscalar constant metric suffices on this box. For $P=\operatorname{diag}(p,1)$ and $0<\lambda<1$, Sylvester's criterion gives strict contraction on $|x_1|\le B$ when $p>B^2/(1-\lambda)^2$. At the exact decimal rate $\lambda=\polyConstRate$ and $B=\polyConstBox$, this threshold is $\polyConstDiagonalThreshold$, so $P_{\mathrm{pol}}=\operatorname{diag}(\polyConstPFirst,\polyConstPSecond)$ is a simple integer choice, giving
\[
C_{P_{\mathrm{pol}}}(x_1)=
\begin{pmatrix}-\polyConstCFirst & \polyConstCross x_1\\
\polyConstCross x_1 & -\polyConstCSecond\end{pmatrix}.
\]
We choose the raw proof margin $\eta=\polyConstRawMargin$ independently of the normalized learner margin $\varepsilon_0$. The diagonal-choice criterion gives strict contraction; shifted endpoint checks establish this margin. The leading principal minors of $-(C_{P_{\mathrm{pol}}}+\eta I)$ at both endpoints $x_1=\pm\polyConstBox$ are $\polyConstShiftLeading$ and $\polyConstShiftDet$, respectively. They are positive, and affine dependence on $x_1$ proves $C_{P_{\mathrm{pol}}}\preceq-\eta I$ throughout $\X$. Also $I\preceq P_{\mathrm{pol}}\preceq\polyConstMetricUpper I$, and $\tilde g\le-\polyConstNormalizedMargin$. The reproduction proves these margins also at the stored binary64 learner rate. This fixed post hoc candidate involves no training, optimized search, or matched timing comparison. Its endpoint proof differs from the learned-metric grid oracle and does not measure its cost. State dependence is unnecessary at the requested rate; all original learning outcomes and budgets are retained.

The certificates give variational decay along trajectories while they remain in $\X$. The box $[-2,2]^2$ is not forward invariant: a trajectory starting at $(2,2)$ leaves through the top face. Finite trajectory-distance bounds require Remark~\ref{rem:finite}; a convex forward-invariant subset suffices. Section~\ref{sec:extension3d} treats an additional three-dimensional example.

Individual certificate upper bounds are rounded upward; margins and robustness budgets are rounded downward. Verification uses saved full-precision values. The descriptive summaries in Table~\ref{tab:certsummaries}, including the median $b$ below, use rounding to nearest. The main polynomial study uses the parametrization \eqref{eq:class} with $\alpha = \numCertA$, target rate $\lambda = 0.4$, an initial dataset of $\PolyInitialSamples$ uniform samples, Adam as the learner with at most $\PolyTrainSteps$ steps per round, $\numSeeds$ seeds, up to $60$ worst margin violators per round with $5$ Gaussian perturbations each of standard deviation $\sigma=\revPerturbStd$, clipped back into $\X$, coupling factor $\gamma = \numGamma$, and a hinge margin base $\varepsilon_0 = \numEpsZero$. The coarse grid has $\numGridCoarse^2$ points and the operating grid $\numGridOp^2$ points, with covering radius $\tau \approx \numTauOp$. Every output has the certified lower bound $a = \numCertA$. The median of the computed upper bounds $b$ over the certified seeds is $\numGridB$; each output uses its own bound in the overshoot $\sqrt{b/a}$ and Theorem~\ref{thm:rate}. The reproduction report records stage runtimes and the execution environment. Separate timing files for the polynomial, pendulum, and neural studies record learner and oracle times and grid evaluation counts per round, including bound computation in oracle time. Table~\ref{tab:timings} reports the machine, learner and oracle times, and evaluation counts. These measurements depend on the machine and verification grid.\footnote{The oracle's Lipschitz constant is a certified upper bound computed by interval coefficient bounds, with box subdivision, on the polynomial entries of $\nabla_x C_\theta$ for the class \eqref{eq:class}; the implementation is included in the companion reproduction archive.}

The pendulum and neural studies use Gaussian standard deviation $\sigma=\revOtherPerturbStd$. The three baseline studies reset Adam moments at each round and use the budgets in Table~\ref{tab:protocol}. A successful run can terminate before its retraining cap. The numerical outputs record these settings together with the grid, subdivision depth, seed, and per-round trace.

\begin{table}[t]
\caption{Maximum learner budgets and initialization sizes}
\label{tab:protocol}
\centering
\small
\begin{tabular}{lrrrr}
\hline
Study & \shortstack{Retraining\\cap $K$} & \shortstack{Adam steps\\per round} & \shortstack{Learning\\rate} & \shortstack{Initial\\samples} \\
\hline
Polynomial & $\PolyRoundCap$ & $\PolyTrainSteps$ & $\PolyLearningRate$ & $\PolyInitialSamples$ \\
Pendulum & $\PendRoundCap$ & $\PendTrainSteps$ & $\PendLearningRate$ & $\PendInitialSamples$ \\
Neural & $\NeuralRoundCap$ & $\NeuralTrainSteps$ & $\NeuralLearningRate$ & $\NeuralInitialSamples$ \\
\hline
\end{tabular}
\end{table}

\subsection{Stall, coupled-margin divergence, and certification}

\begin{figure}[t]
\centering
\includegraphics[width=0.72\textwidth]{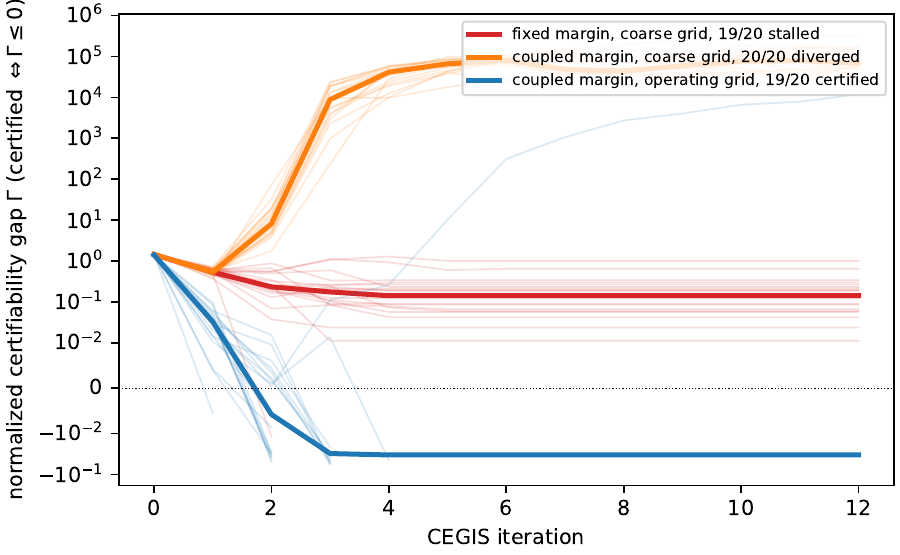}
\caption{Normalized certifiability gap $\Gamma$ per CEGIS iteration over $\numSeeds$ seeds (thin lines) with per-regime medians (bold), on a symlog axis so certification appears as a crossing below zero. All three regimes train on the normalized test. The coarse-grid regimes use fixed margin or coupled margin; the operating-grid regime uses coupled margin. Counts and exceptional seed outcomes are reported in the text. Labels describe finite-budget outcomes rather than exact stationary points or infinite-time divergence.}
\label{fig:traces}
\end{figure}

Figure~\ref{fig:traces} shows the normalized certifiability gap $\Gamma$ of Definition~\ref{def:gap} over $\numSeeds$ seeds for three regimes of the loop, all trained on the normalized test so the gaps share one scale-invariant axis. An uncertified run is labeled diverged when $\max_k L_k>\OutcomeGrowthFactor L_0$. In the polynomial study, the remaining uncertified runs are labeled stalled. These labels describe outcomes within the budgets of Table~\ref{tab:protocol}; they do not prove exact stationary points or infinite-time divergence. With a fixed margin $\varepsilon_0 = \numEpsZero$ on the coarse grid of $\numGridCoarse^2$ points, the loop stalls in $\numStallCount$ of $\numSeeds$ runs. At the stalled candidates, the margin lies below the oracle's requirement $L_h\tau$, so the stall band of Proposition~\ref{lem:stall} is nonempty. Coupling the margin on the same coarse grid raises the next training margin above the current oracle requirement, but the certified $L_h$ is conservative, so the required margin $\gamma L_h\tau$ outgrows the slack achieved by the learner, and the loop diverges in $\numDivCount$ of $\numSeeds$ runs, increasing the certified Lipschitz bound for the normalized test by up to a factor $\numDivGrowth$. On the operating grid of $\numGridOp^2$ points, where the learner can achieve slack above $\gamma L_h\tau$, Algorithm~\ref{alg:cegis} certifies in $\numCertCount$ of $\numSeeds$ runs with a median of $\numCertMedIters$ retraining rounds. \revLegacySeedsSummary

\begin{figure}[t]
\centering
\includegraphics[width=0.95\textwidth]{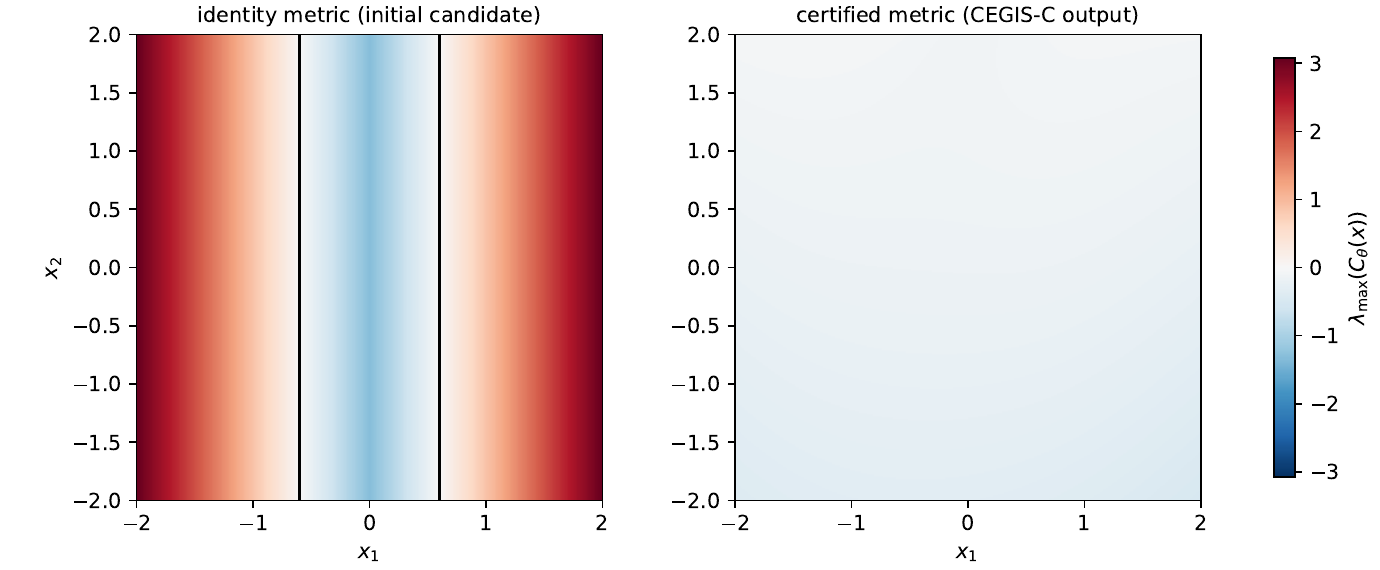}
\caption{$\lmax(C_\theta(x))$ over $\X$ for the initial identity-scaled candidate (left, zero contour in black at $|x_1| = 0.6$) and the certified CEGIS-C output (right, uniformly negative).}
\label{fig:heatmaps}
\end{figure}

The scale chase of Corollary~\ref{cor:chase} is distinct from the growth of the shape-dependent normalized Lipschitz bound observed here. Figure~\ref{fig:heatmaps} shows the condition heatmaps: the identity-scaled candidate violates on $|x_1| > 0.6$ as predicted, and a certified output is uniformly negative.

\subsection{Controlled ablations and bounded learners}
\label{sec:ablation}

The new ablation crosses raw versus normalized tests with fixed versus coupled margins on the same operating grid, with matched seeds, initial parameters, initial samples, and optimizer budgets. These are complete loss-and-oracle configurations: the raw and normalized margins have different units even when their numerical base values agree, and their certified bounding rules have different conservatism. The comparison does not isolate optimization conditioning from verification-bound quality. Table~\ref{tab:ablation} gives every outcome, including uncertified and numerical-failure cases. A second, prespecified comparison constrains coupled-margin training to a parameter ball, using projection after every Adam update and a radius chosen from the common initialization before observing outcomes. \revBoundedRadius{} This is a coupled-margin subset, rather than a full factorial study of all interactions with boundedness. The ball bounds the parameter sequence but neither proves zero empirical loss nor realizes the idealized learner in Proposition~\ref{prop:termination}. It also does not remove the finite-budget relative-growth label: the bounded experiments can still satisfy $\max_kL_k>\OutcomeGrowthFactor L_0$. A uniform finite bound on $L_h$ can be much larger than this multiple of its initial value, so that label must not be read as unbounded parameter growth. A separate bounded coarse-grid regime tests whether a parameter constraint controls the observed growth without assuming that this grid margin is feasible. \revAblationSummary

The matched operating-grid results do not show a certification-count benefit from normalization or coupling. They support these choices as safeguards against the stated mechanisms, rather than necessary improvements on this benchmark. In the raw runs, the coupled margin never exceeds the base margin, so fixed and coupled training follow the same trajectory. This also limits the extent to which that comparison tests coupling.

Sensitivity checks vary the coupling factor and returned-point count one at a time. Table~\ref{tab:sensitivity} records their smaller, separately specified coarse-grid budgets. \revSensitivitySettings{} They are short-horizon diagnostics, not estimates of the operating-grid certification rate. A fixed-margin pendulum comparison is reported under its stated budget as an additional check rather than inferred from the polynomial case.

To connect Corollary~\ref{cor:chase} to a concrete family, we use $M_c=cM_0$, $c\ge1$, for a uniformly contracting polynomial metric $M_0$ that remains rejected on a coarse grid. This family is closed under upward scaling and preserves the positive lower metric bound at $c=1$. At each step, the next scale satisfies the corollary's sufficient inequality for the coupled margin $\varepsilon_k=\gamma L_k\tau$ supplied by the previous candidate. Directed bounds check uniform raw negativity, satisfaction of that training margin, and a strictly positive coarse-grid certifiability gap. The finite plotted prefix illustrates the proof construction; it is not evidence of an infinite numerical run. The normalized test and its exact bound remain unchanged. It is not an observed Adam trajectory. Normalization prevents scaling from improving the test but cannot certify a frozen shape that remains rejected. \revScaleChaseSummary

\IfFileExists{fig_ablation.pdf}{%
\begin{figure}[t]\centering
\includegraphics[width=0.85\textwidth]{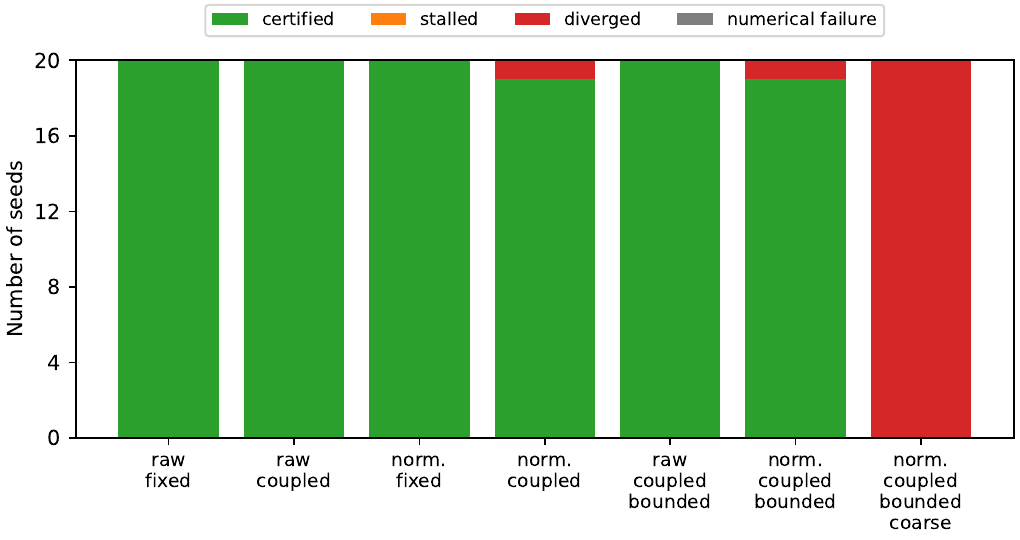}
\caption{Matched-seed ablation outcomes. The primary comparison crosses raw/normalized tests and fixed/coupled margins on one grid. Bounded coupled-margin variants and the coarse-grid variant are separate comparisons. Counts describe the recorded finite budgets.}
\label{fig:ablation}
\end{figure}}{}

\IfFileExists{fig_scale_chase.pdf}{%
\begin{figure}[t]\centering
\includegraphics[width=0.85\textwidth]{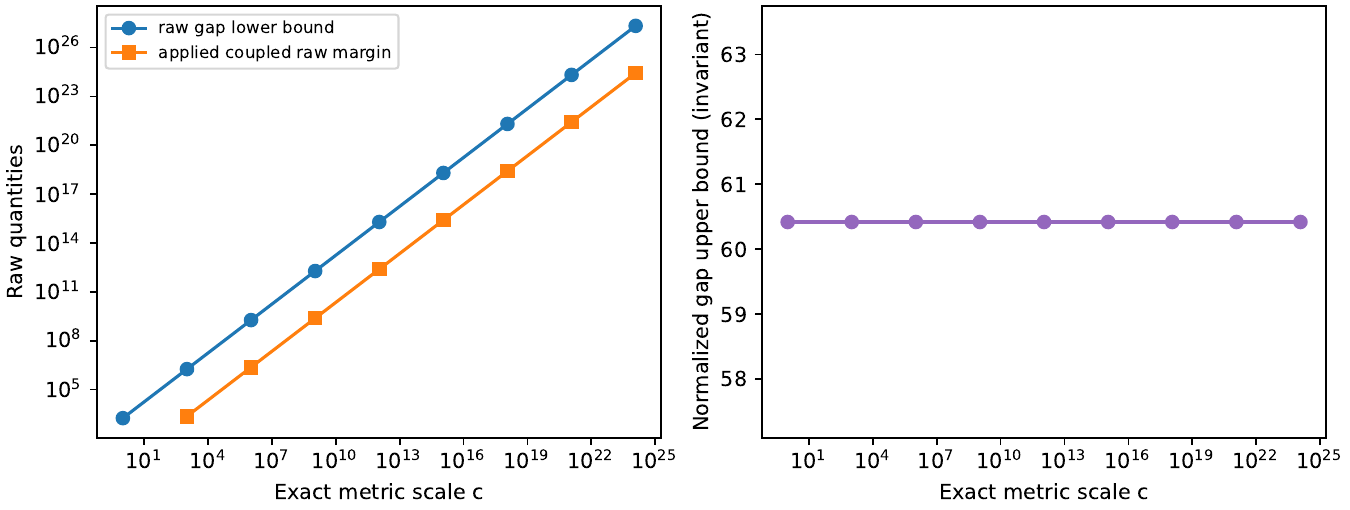}
\caption{Constructive scale chase in the family $M_c=cM_0$. Increasing raw scale can meet the next training requirement while the raw gap stays positive; the normalized gap is unchanged. Every plotted bound comes from directed verification. This illustrates Corollary~\ref{cor:chase}, not a gradient optimizer's behavior.}
\label{fig:scalechase}
\end{figure}}{}

\noindent\textit{Fixed grids and refinement.}\ We hold the grid fixed within each baseline run to isolate training changes from changes in the verification requirement. For deployment, a budgeted alternative is to freeze a rejected metric and refine the grid, or invoke an adaptive box oracle, before retraining. Table~\ref{tab:refinement} tests axis-grid refinement on selected previously certified metrics, with no learner updates. Each failed recheck is inconclusive; each positive answer again requires the complete outward-rounded grid check. This diagnostic is conditional on the selected metrics and does not estimate a coarse-grid training success rate or establish termination of a refining learner-verifier loop.

\noindent\textit{Direct gap training.} A possible alternative is to minimize
\[
\frac1{|\D|}\sum_{x\in\D}\max\{0,\tilde g_\theta(x)+\gamma L_h(\theta)\tau\}.
\]
This penalizes variation of the bounding function during a training step. It requires a differentiable or subgradient-compatible surrogate of the certified coefficient bound, followed by the exact verification check. We keep the external margin update in the ablation to compare the prescribed test, margin-rule, and parameter-constraint configurations under shared budgets. Direct gap training is not evaluated here, and no optimality claim is made for external coupling.

\subsection{The cost of the grid oracle}

By Lemma~\ref{lem:gridsuff} the oracle certifies at resolution $\tau$ whenever the achieved normalized slack exceeds $L_h \tau$, and certification in turn implies a nonnegative uniform slack; the reverse equivalence fails, since grid values can clear the margin with true slack slightly below $L_h\tau$. By Lemma~\ref{lem:scale} the ratio $\mathrm{slack}/L_h$ is scale-invariant when the bounding rule respects scaling. Across the $\numCertCount$ certified candidates we estimate slack as the negative maximum of the normalized test on a separate dense grid. This sampled slack is an upper bound on the true uniform slack, not a certified lower bound. Its median ratio to the certified Lipschitz bound is $\numGridMedianRatio$, with range $[\numGridMinRatio, \numGridMaxRatio]$, and assuming that this measured two-dimensional ratio stays fixed as $n$ changes gives the following illustrative grid-count extrapolation: $\numGridPtsTwo$ points at $n = 2$, $\numGridPtsFour$ at $n = 4$, and $\numGridPtsSix$ at $n = 6$. These counts are estimates, not certified minimum grid requirements or experiments in higher dimensions. At a budget of $10^9$ evaluations the extrapolation crosses the budget between $n = 3$ and $n = 4$. The counts describe the stored uniform grid. Interval grouping can reduce the cost of proving its inequalities, while the nominal rejection screen still evaluates the full grid in this implementation. The assumption is untested in higher dimensions; the ratio can change with the dynamics, metric, and enclosure quality. The extrapolation illustrates the cost of uniform coverage. Adaptive interval subdivision can exploit local slack, but can also require exponentially many boxes in the worst case. Section~\ref{sec:extension3d} compares box-wise verification with this baseline.

\subsection{Residual robustness of the certified metric}

\begin{figure}[t]
\centering
\includegraphics[width=0.62\textwidth]{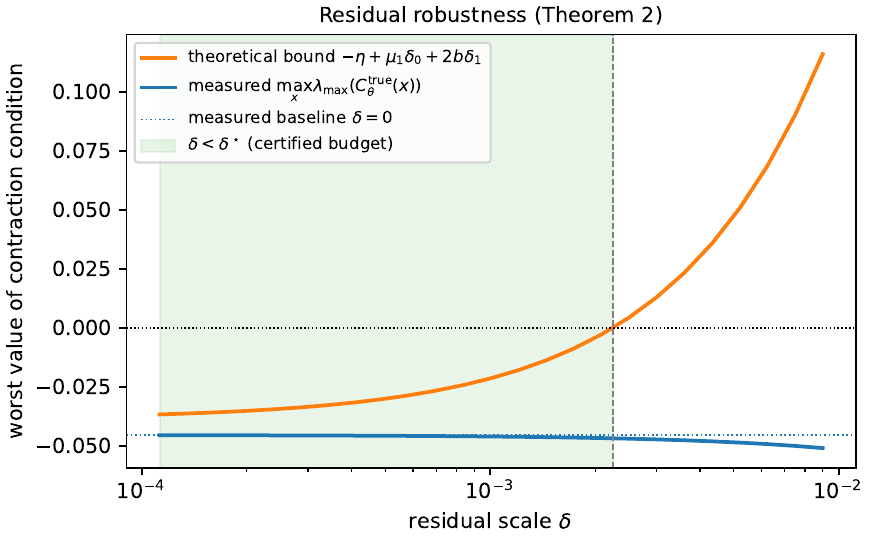}
\caption{Residual robustness (Theorem~\ref{thm:residual}): measured worst value of the true contraction condition versus the theoretical bound as the residual scale $\delta$ grows. The shaded region is the certified budget $\delta < \delta^\star$; its boundary is marked explicitly. The horizontal reference is the measured zero-residual value. Measurements outside the shaded region are diagnostic.}
\label{fig:residual}
\end{figure}

We exercise Theorem~\ref{thm:residual} on the benchmark. A certified candidate attains a certified lower raw margin $\eta = \numEta$ on $\X$. This value comes from a separate directed-rounded raw-test grid check, subtracting the certified $L_g\tau$ from the negative upper grid maximum; it is not the normalized conversion $na\underline s$. The raw-grid check uses seed $\revResidualSeed$, $\revResidualGridAxis$ points per axis, and coefficient subdivision depth $\revResidualDepth$; its covering radius and metric hash are recorded in the output. The certified bounds $b \le \numBupper$ and $\mu_1 \le \numMuOne$ use Lemma~\ref{lem:certconst}(iii)--(iv) with the recorded subdivision depth. We perturb the dynamics with the polynomial residual $d(x) = \delta \cdot 0.25\,(x_2^2,\; x_1 x_2)$, for which certified upper bounds on the residual constants are $\delta_0 = \numCdZero\,\delta$ and $\delta_1 = \numCdOne\,\delta$, so the theorem guarantees contraction of the true system for $\delta < \delta^\star = \numDeltaStar$. Figure~\ref{fig:residual} compares, across a sweep of $\delta$, the measured worst value of the true contraction condition on the verification grid against the theoretical bound $-\eta + \mu_1\delta_0 + 2b\delta_1$. The measured curve stays below the bound throughout. The largest tested residual scale is $\numDeltaEmpRatio$ times the predicted budget, and all sampled condition values remain negative through that endpoint. This observation does not locate a failure threshold or certify the region beyond $\delta^\star$. The gap between the curves illustrates the conservatism of the norm-based bound on this benchmark.

\subsection{A benchmark with trigonometric dynamics}

To test the framework beyond the polynomial class, we apply it to the damped pendulum $f(x) = (x_2,\ -\sin x_1 - x_2)$ on $\X = [-\numPendBox, \numPendBox]^2$, whose dynamics are not polynomial in the original state coordinates. An exact polynomial lift uses $s=\sin x_1$, $c=\cos x_1$, and $s^2+c^2=1$ \cite{tedrake2026}. This permits SOS analysis in lifted coordinates; translating the metric class and the original domain requires care. Our oracle instead encloses the trigonometric terms directly in the original coordinates. The metric class is unchanged, so $M_\theta$ and its derivatives stay polynomial and only the contraction matrix $C_\theta$ acquires the atoms $\sin x_1$ and $\cos x_1$. We certify it with the same interval bounds by carrying $C_\theta$ and its partials as polynomials in $(x_1, x_2, \sin x_1, \cos x_1)$, closed under the derivations $\partial_{x_1}: \sin x_1 \mapsto \cos x_1$, $\cos x_1 \mapsto -\sin x_1$, and bounding $\sin x_1$ and $\cos x_1$ as atoms of magnitude at most one, tightened on each subbox by directed center evaluation plus a Lipschitz remainder bounded by the coordinate half-width. The analytic enclosure needs no polynomial-approximation remainder. Outward rounding also covers the coefficient algebra and trigonometric enclosures. Dense sampling checks implementation consistency; the interval construction supplies the enclosure guarantee.

\begin{figure}[t]
\centering
\includegraphics[width=0.95\textwidth]{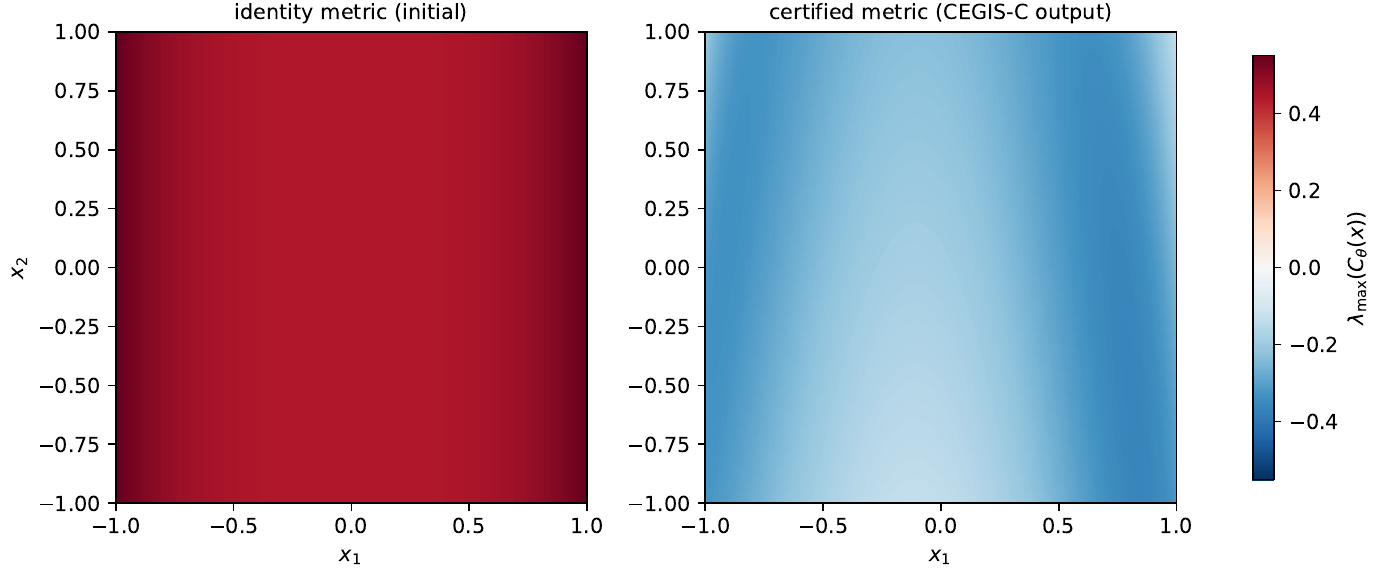}
\caption{Non-polynomial benchmark (damped pendulum): $\lmax(C_\theta(x))$ for the identity metric (left, positive throughout) and the certified CEGIS-C output (right, uniformly negative), certified by the trig-polynomial interval oracle.}
\label{fig:pendulum}
\end{figure}

Unlike the constructed polynomial benchmark, this study does not assume a known certifiable metric in the chosen class. It does not establish infeasibility at other target rates. At rate $\lambda = \numPendRate$ the loop certifies in $\numPendCert$ of $\numPendSeeds$ seeds with a median of $\numPendMedIters$ retraining round and diverges in $\numPendDiv$, showing both certification and divergence under the normalized coupled-margin loop. This study tests certification beyond polynomial dynamics. Its fixed-margin and parameter sensitivity behavior is not inferred from the polynomial benchmark. A certified candidate attains certified constants $L_h = \numPendL$, a lower bound $\numPendSlack$ on grid slack, and $b \le \numPendB$, and it passes the oracle on the operating grid of $\numPendGrid^2$. The grid slack is the negative maximum at grid points; subtracting $L_h\tau$ gives a certified uniform normalized margin in real arithmetic. Figure~\ref{fig:pendulum} shows the condition heatmaps: the identity metric is positive throughout, and the certified output is uniformly negative.

\subsection{A neural metric parametrization}
\label{sec:neural}

We replace the linear-in-features Cholesky factor with a one-hidden-layer $\tanh$ network of $\numNeuralUnits$ units. The three-dimensional output $\ell(x)=W_2\tanh(W_1x+b_1)+b_2$ supplies $(\ell_{11},\ell_{21},\ell_{22})$ of a lower-triangular factor $L(x)$, and $M_\theta=\alpha I+L^\top L$. Here $W_1\in\R^{\numNeuralUnits\times2}$ and $W_2\in\R^{3\times\numNeuralUnits}$, on the polynomial benchmark, using the same normalized, coupled-margin principle. The neural study also uses its own initialization, optimizer settings, and finer verification grid, as recorded in the implementation and reproduction outputs; it is not a controlled comparison of expressiveness alone. The metric is no longer polynomial, so the coefficient-bound oracle does not apply. We certify with outward-rounded interval arithmetic that propagates enclosures through the exact computation of $C_\theta$ and its partials, enclosing $\tanh$ and its first two derivatives by interval operations; the analytic enclosure follows from interval arithmetic. Outward rounding encloses parameter products, nonlinear terms, matrix entries, and grid evaluations. Dense sampling and finite-difference checks test the implementation; positive verification uses interval endpoints.

\begin{figure}[t]
\centering
\includegraphics[width=0.95\textwidth]{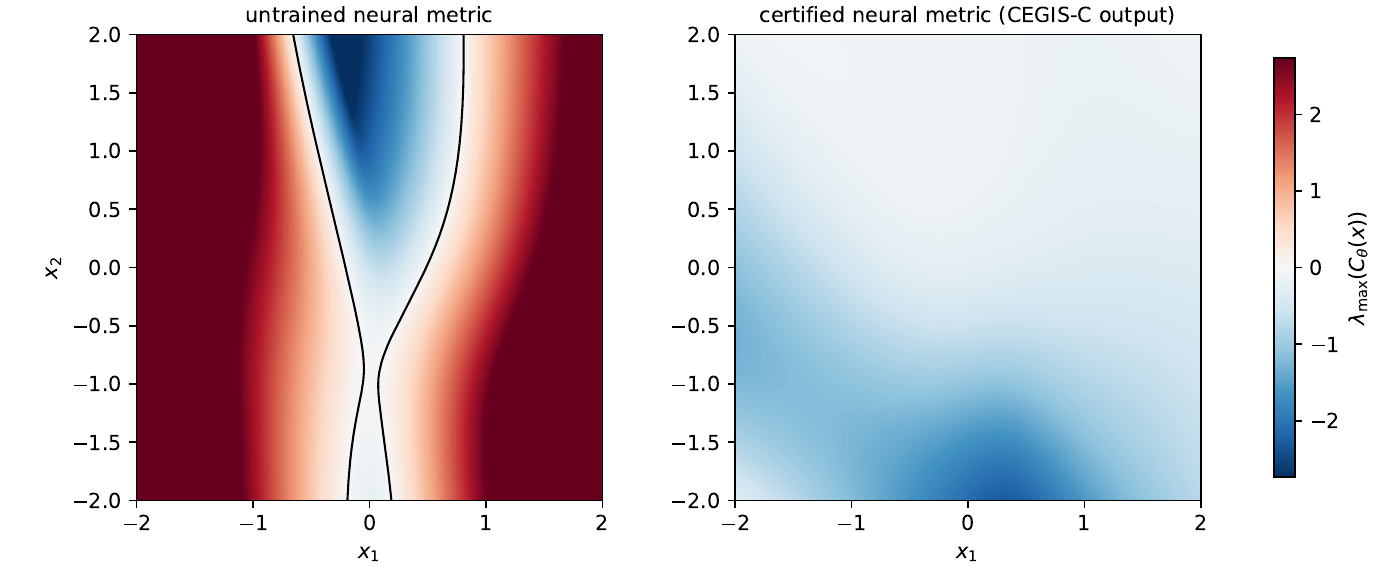}
\caption{Neural metric ($\numNeuralUnits$-unit $\tanh$ network): $\lmax(C_\theta(x))$ for the untrained network (left) and a certified CEGIS-C output (right, uniformly negative), certified by interval bound propagation through the network.}
\label{fig:neural}
\end{figure}

The neural parametrization drives the normalized condition below zero in successful runs. In this configuration, its certified bounds and achieved slack lead to certification on a finer grid and more frequent finite-budget divergence than in the linear-feature study. The learned functions, initialization, optimizer, and oracle all change, so the comparison does not isolate interval-bound conservatism or representational expressiveness. A larger certified Lipschitz bound can reflect both actual metric variation and enclosure overestimation. On the operating grid of $\numNeuralOpGrid^2$ points the loop certifies $\numNeuralCert$ of $\numNeuralSeeds$ seeds with a median of $\numNeuralMedIters$ retraining round and diverges in $\numNeuralDiv$; of the certified metrics only $\numNeuralCoarse$ still passes the oracle at the linear class's grid of $\numNeuralCoarseGrid^2$. A certified candidate attains $L_h=\numNeuralL$, a lower bound $\numNeuralSlack$ on normalized grid slack, and global metric bounds $a=\numNeuralA$ and $b\le\numNeuralB$. Its certified uniform normalized slack is at least $\numNeuralUniformSlack$. The saved outputs include the raw margin, improved rate, and overshoot computed with directed bounds. The coarse-grid recheck tests these already learned metrics; it does not measure the success rate of training the neural loop on the coarse grid. The constant certificate in Section~\ref{sec:polynomialbenchmark} already meets the requested rate on this domain. These neural experiments therefore exercise the synthesis workflow and grid-cost constraint, without demonstrating a need for neural or state-dependent metrics. Figure~\ref{fig:neural} shows a certified condition, uniformly negative.

\subsection{Adaptive interval verification and a three-dimensional loop}
\label{sec:extension3d}

The adaptive polynomial oracle partitions $\X$ into boxes. An exact rational conversion puts each polynomial entry of $C_\theta$ into tensor-product Bernstein form on the initial box. The convex-hull property bounds an entry by the smallest and largest of its Bernstein coefficients. Coefficient conversion is rounded outward; subsequent bisections propagate outward coefficient intervals by de Casteljau subdivision at the exact split fraction defined by the stored endpoints. To prove a prescribed raw margin $\rho>0$, an interval $LDL^\top$ calculation checks $B_\theta=-C_\theta-\rho I$ with strictly positive lower pivot bounds. If direct factorization is inconclusive, a nominal center calculation may select a fixed upper-triangular matrix $R$ with finite stored entries and nonzero diagonal. The stored matrix is interpreted exactly and is invertible by triangularity. The verifier encloses $R^\top B_\theta R$ outward and repeats the interval pivot test; the nominal factorization itself never supplies a certificate. A passing leaf therefore proves $C_\theta(x)\prec-\rho I$ throughout its box. An unresolved leaf is bisected along its coefficient-weighted longest edge. The coefficient-derived weights and tie rule select a partition; they do not supply any positivity guarantee. The oracle certifies only when all leaves pass; depth, box, resolution, or arithmetic limits return inconclusive. Nominal sampled eigenvalues select training points and may screen a candidate out, but never produce a positive answer. This oracle does not require a Lipschitz constant or a uniform grid. Its worst-case cost can still grow exponentially.

\begin{proposition}[Adaptive oracle soundness]
\label{prop:adaptive}
If the adaptive oracle certifies all leaves at raw margin $\rho>0$, then $C_\theta(x)\prec-\rho I$ on $\X$.
\end{proposition}
\begin{proof}
The Bernstein convex-hull bound and outward subdivision enclose each matrix entry on every leaf. For each fixed point, interval $LDL^\top$ operations enclose the exact factorization of $B_\theta(x)$ or of $R^\top B_\theta(x)R$ for a fixed invertible stored $R$. Strictly positive lower pivot bounds prove positive definiteness of the factored matrix. In the second case, Sylvester's law of inertia gives $B_\theta(x)\succ0$ by invertible congruence. The accepted leaves cover $\X$, giving the claim. Unresolved or unprocessed leaves prevent certification.
\end{proof}

We test the coupled polynomial system
\[
f(x)=(-x_1,\,-x_2+x_1^2,\,-x_3+x_1x_2-x_1^3)
\quad\text{on }[-1,1]^3.
\]
The transformation $\varphi=(x_1,x_2+x_1^2,x_3+x_1x_2)$ obeys $J_\varphi f=-\varphi$ and has a unit lower-triangular Jacobian. Thus $J_\varphi^\top J_\varphi$ is an independent reference contraction metric; the reproduction checks the polynomial identity exactly. The learner does not use this metric for initialization or training. It searches all lower-triangular factors linear in $(1,x_1,x_2,x_3)$, giving 24 parameters, from seeded perturbations of the identity factor. \revExtensionSettings{}
The settings, oracle budgets, and outcomes appear in Table~\ref{tab:extensions} and the saved traces. This is a new constructed benchmark, rather than an SOS comparison or evidence of performance on arbitrary higher-dimensional systems. \revExtensionSummary

The same adaptive oracle also rechecks frozen planar metrics selected from the original certified outputs. This holds the metric and domain fixed. The grid oracle proves the base-rate condition through a normalized Lipschitz margin, while the adaptive oracle requests a prescribed positive raw margin. Grid points and box operations are different units of work, so the table reports each separately with wall times. The comparison does not by itself establish a universal speedup or equal-difficulty certification tasks.

\IfFileExists{fig_extension3d.pdf}{%
\begin{figure}[t]\centering
\includegraphics[width=0.85\textwidth]{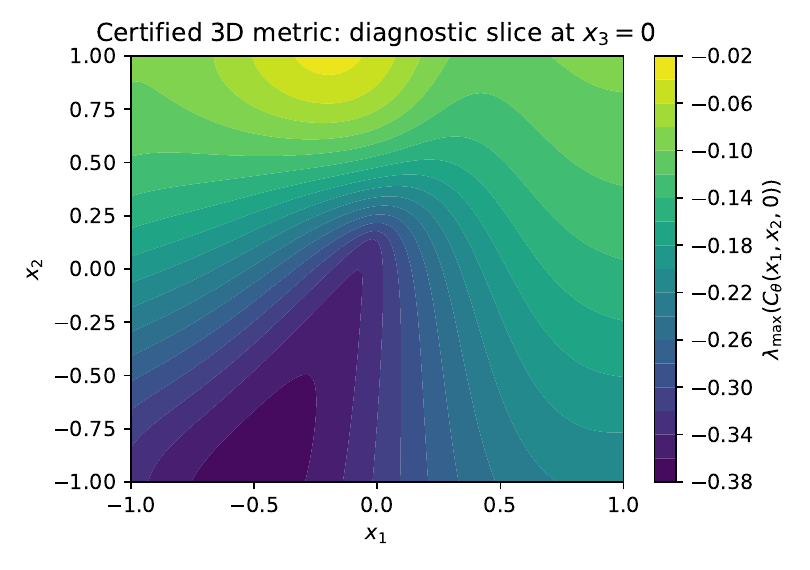}
\caption{Diagnostic slice $\lmax(C_\theta(x_1,x_2,0))$ for the first certified three-dimensional seed. The slice does not prove negativity throughout the volume; the full-box certificate comes from the adaptive partition with exact polynomial construction, outward Bernstein subdivision, and interval matrix checks on every accepted leaf.}
\label{fig:extension3d}
\end{figure}}{}

\subsection{Transfer from a trajectory-fitted surrogate}
\label{sec:surrogate}

We fit a quadratic polynomial vector field to trajectory observations of the planar benchmark. Initial states and observation times are fixed by the recorded seed and sampling protocol. Central differences of noiseless position observations provide derivative targets; ordinary least squares fits all monomials of total degree at most two. The exact true vector field is not supplied to the regression or metric learner. The surrogate coefficients are then treated as exact stored binary64 values for verification. A separate adaptive loop learns and certifies a metric for the fitted model. \revSurrogateSettings{}

For this known-truth benchmark only, we form the exact polynomial residual $d=f_{\mathrm{true}}-f_{\mathrm{fit}}$ after fitting. Corollary~\ref{cor:polyresidual} certifies its magnitude and Jacobian bounds on the whole verification box, independently of the sampled regression error. The metric bounds and metric-derivative constant then give the transfer test in Theorem~\ref{thm:residual}. Table~\ref{tab:extensions} reports fitting settings, global residual bounds, consumed margin, and the resulting true-system rate when the strict inequality passes. This experiment exercises a fitted surrogate with discretization error; it does not infer global error bounds for an unknown system or cover noisy data. \revSurrogateSummary

\IfFileExists{fig_surrogate.pdf}{%
\begin{figure}[t]\centering
\includegraphics[width=0.7\textwidth]{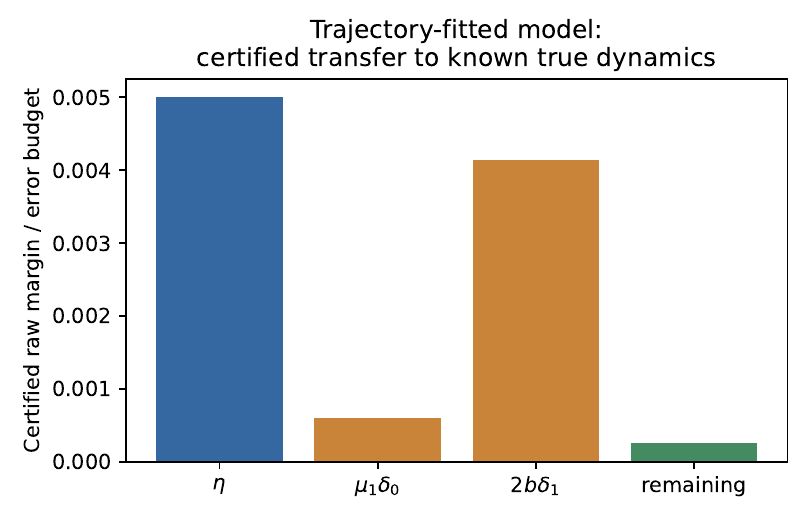}
\caption{Certified margin and model-error budget for the trajectory-fitted surrogate. The terms $\mu_1\delta_0$ and $2b\delta_1$ are outward upper bounds; any displayed positive remainder is rounded downward. Global residual certification uses the independently known polynomial truth.}
\label{fig:surrogate}
\end{figure}}{}

\subsection{A physical swing model with a matched PDI comparison}
\label{sec:application}

We use the two-machine power-system model in Example~3 of Fitzsimmons and Liu~\cite{fitzsimmons2024}. Its states are relative rotor-angle deviation $\delta$ and angular-speed deviation $\omega$:
\[
\dot\delta=\omega,\qquad
\dot\omega=-\tfrac12\omega-\bigl(\sin(\delta+\pi/3)-\sin(\pi/3)\bigr).
\]
This is a physical-model benchmark rather than operational power-grid data. We use the published dynamics without fitting them or constructing them from a known contraction metric. The fixed protocol tests every combination of $B\in\{\appBSmall,\appBMedium,\appBLarge\}$, $\appSeeds$ paired seeds, and three learners on $\X_B=[-B,B]^2$, for $\appTotalRuns$ runs. The requested rate is $\appRate$. The largest box extends beyond the angle where the restoring-force derivative changes sign; all boxes remain short of the unstable equilibrium at $(\pi/3,0)$. These domains and budgets were fixed before the comparison and are reported even when certification fails.

\noindent\textit{Matched objectives and feedback control.}
Every learner uses the same $\appUnits$-unit tanh factor with $\appParameters$ parameters, metric floor $\appFloor$, initial parameters, and $\appSamples$ uniform collocation points. Write $R_\theta(x)=C_\theta(x)+qI$, with the common raw training margin $q=\appTrainMargin$. The CEGIS learner minimizes
\[
\frac1{|\mathcal B|}\sum_{x\in\mathcal B}
\max\left\{0,\frac{\lmax(C_\theta(x))+q}{\tr M_\theta(x)}\right\}.
\]
After a rejected check, it accumulates at most $\appReturned$ diagnostic points. This adaptive-oracle experiment uses a fixed raw slack target rather than the grid-dependent coupled margin in Algorithm~\ref{alg:cegis}. The PDI baseline independently implements the spectral loss of \cite[Eq.~(5)]{fitzsimmons2024}:
\[
\frac1{2|\mathcal B|}\sum_{x\in\mathcal B}\sum_{j=1}^2
\max\{0,\lambda_j(R_\theta(x))\}^2.
\]
Its positivity penalty vanishes under the shared factor construction with $S=\appFloor I$. It trains on the fixed initial collocations. Both losses have the same pointwise zero set, $R_\theta\preceq0$. A third learner uses the normalized hinge loss on those same fixed collocations. This control separates diagnostic augmentation from the change of loss. It is an internal ablation, not another published baseline.

The common budget is $\appRounds$ blocks of $\appSteps$ Adam updates, learning rate $\appLearningRate$, and batch size $\appBatch$, with moments reset at every block. The initial candidate and each block's output receive the same verification call. Each learner stops at its first certificate; otherwise it uses at most $\appTotalSteps$ updates. CEGIS draws seeded batches without replacement once its accumulated dataset exceeds the batch size. Method order reverses on odd seeds, and methods run serially. Tables~\ref{tab:application} and~\ref{tab:applicationpaired} report all-run medians and paired outcomes; equal update caps are not equal wall-clock budgets.

The comparison changes the original PDI implementation's architecture, positivity parametrization, prescribed matrix target, collocation budget, invariant-domain construction, and verifier. In particular, $R_\theta$ includes the shared requested rate and raw margin, whereas the original examples use prescribed $Q=S=I$. We implement the published equations independently and record their mapping to the code. This is a matched learning-objective comparison, not a reproduction or ranking of the full LyZNet/SMT method.

\noindent\textit{Common numerical certification.}
The verifier rewrites the restoring force exactly as
$\tfrac12\sin\delta+\tfrac{\sqrt3}{2}(\cos\delta-1)$.
Directed square-root and MPFR trigonometric enclosures retain the exact physical phase $\pi/3$. The network and its analytic spatial derivatives give an outward interval enclosure of each contraction-matrix entry. For a symmetric two-by-two interval matrix, its largest possible eigenvalue is bounded by the upper diagonal endpoints and the largest absolute off-diagonal endpoint. Adaptive binary subdivision accepts a candidate only when every leaf proves $C_\theta\preceq-\appVerifyMargin I$. Each call allows at most $\appBoxBudget$ evaluated boxes and depth $\appDepthBudget$. A $\appScreen\times\appScreen$ nominal grid can reject a candidate but cannot certify it. The full binary leaf cover and every positive matrix inequality are replayed independently of training. Exhausted budgets and nominal-screen rejection are inconclusive outcomes, not proofs that the dynamics lack a contraction metric.

\noindent\textit{An invariant region for synchronization.}
The boxes themselves are not forward invariant. Define
\[
U(\delta)=\tfrac12(1-\cos\delta)+\tfrac{\sqrt3}{2}(\sin\delta-\delta),
\qquad E(\delta,\omega)=\tfrac12\omega^2+U(\delta).
\]
Then $\dot E=-\omega^2/2$. For the chosen boxes, $\delta U'(\delta)>0$ away from zero because $B<\pi/3$ and
\[
U'(\delta)=2\sin(\delta/2)\cos(\pi/3+\delta/2).
\]
Consequently $U\ge0$ in the box. Its boundary energy is at least
\[
\min\{U(-B),U(B),B^2/2\}.
\]
We compute a downward-rounded level $\rho_B$ below nine tenths of a certified lower bound on this minimum. The set
\[
\Omega_B=\{(\delta,\omega)\in[-B,B]^2:E(\delta,\omega)\le\rho_B\}
\]
lies strictly inside the box and is forward invariant. The certified levels are $\appEnergySmall$, $\appEnergyMedium$, and $\appEnergyLarge$ in increasing order of $B$. These are levels of the in-box sublevel set, not of the unrestricted tilted potential. The set is star-shaped about the synchronized equilibrium, so radial paths flow within it. A successful contraction certificate therefore implies exponential convergence to that equilibrium for every initial state in $\Omega_B$, with the certified metric prefactor. For arbitrary pairs, the connector through the origin has initial length $\|x_0\|+\|y_0\|$. For the two smaller boxes, $U''>0$, so $\Omega_B$ is convex and Remark~\ref{rem:finite} gives the usual initial Euclidean-distance bound. An energy region without a successful metric certificate does not acquire the requested exponential-rate guarantee.

\noindent\textit{Analytical constant-metric baseline.}
After inspecting the neural comparison, we checked the fixed matrix
\[
P=\begin{pmatrix}1&1/2\\1/2&2\end{pmatrix},
\qquad k(\delta)=\cos(\delta+\pi/3).
\]
At the exact decimal rate $\lambda=1/20$, its contraction matrix is
\[
C_P(k)=A^\top P+PA+2\lambda P
=\begin{pmatrix}1/10-k&4/5-2k\\4/5-2k&-4/5\end{pmatrix}.
\]
Directed MPFR bounds place $k$ in $[\appConstKLower,\appConstKUpper]$ on the smallest box. For the requested raw margin $\eta=1/10000$, exact rational checks show that $-(C_P+\eta I)$ has positive leading principal minors at both endpoints. Its affine dependence on $k$ then proves $C_P\preceq-\eta I$ throughout the box. The reproduction records these endpoint proofs, positivity bounds for $P$, and an additional exact check at the actual stored binary64 rate and margin used by the neural verifier. Thus the baseline meets the same numerical targets without relying on decimal/binary64 identification.

Table~\ref{tab:applicationconstant} reports the fixed matrix on every application box. The small-box certificate supports the same invariant-region interpretation on $\Omega_{\appBSmall}$. On each larger box, a directed bound proves $[C_P(k(B))]_{11}>0$, so this matrix fails the whole-box contraction condition. These witnesses lie outside the interior energy sublevels; they do not rule out contraction there or other constant metrics on the box. This is a post hoc analytical candidate, with no learning, random seeds, or optimized search over constant metrics. It is not a matched runtime comparison. Its success establishes that a neural or state-dependent metric is unnecessary for the small-box result, so the learned certificates demonstrate a synthesis-and-verification workflow rather than a need for neural expressiveness.

\noindent\textit{Results and limits.}
In increasing box order, the CEGIS hinge learner certifies $\appCegisSmall$, $\appCegisMedium$, and $\appCegisLarge$ of $\appSeeds$ seeds; the PDI adaptation certifies $\appPdiSmall$, $\appPdiMedium$, and $\appPdiLarge$; the fixed-data hinge control certifies $\appFixedSmall$, $\appFixedMedium$, and $\appFixedLarge$. Table~\ref{tab:applicationpaired} preserves pairing rather than treating all domain--seed combinations as independent trials. The fixed-data hinge control matches the primary CEGIS certification counts, so the primary budget does not demonstrate a diagnostic-feedback advantage. The small-box difference from the PDI adaptation is limited to $\appPdiDiscordantSmall$ discordant seeds. These are descriptive outcomes under one shared optimizer setting and finite verification budgets. The comparison does not tune each loss separately or establish superiority to the original published method, its SMT verifier, or a controller-synthesis method. It adds a physical application and a controlled comparator while retaining the distinction between learned feasibility and numerical certifiability.

After inspecting preliminary primary outcomes, we specified an exploratory frozen-candidate check. Every final candidate, including primary successes and all failures, receives the same enlarged budget of $\appRefinementBudget$ boxes. Parameters, domain, rate, requested raw margin, screening grid, and maximum depth remain unchanged. The check performs no further learning and does not replace the primary counts. Table~\ref{tab:applicationrefinement} reports additional certifications and the added verification cost for every arm. In increasing box order, the enlarged-budget counts are $\appRefCegisSmall$, $\appRefCegisMedium$, and $\appRefCegisLarge$ for CEGIS, $\appRefPdiSmall$, $\appRefPdiMedium$, and $\appRefPdiLarge$ for the PDI adaptation, and $\appRefFixedSmall$, $\appRefFixedMedium$, and $\appRefFixedLarge$ for the fixed-data hinge control. The enlarged budget adds $\appRefCegisMedium$ middle-box CEGIS certificates and no additional certificates for the fixed-data learners. Most remaining middle-box CEGIS candidates exhaust the box budget, whereas the fixed-data candidates are rejected by the nominal margin screen. The added certificates show that the primary budget concealed valid frozen candidates in these cases; this small exploratory difference does not establish a general advantage of feedback. Neither outcome establishes mathematical infeasibility. This secondary analysis tests numerical certifiability of frozen metrics. Its post hoc motivation and common enlarged budget preclude interpreting it as a prespecified general performance claim.

\begin{figure}[t]
\centering
\includegraphics[width=\linewidth]{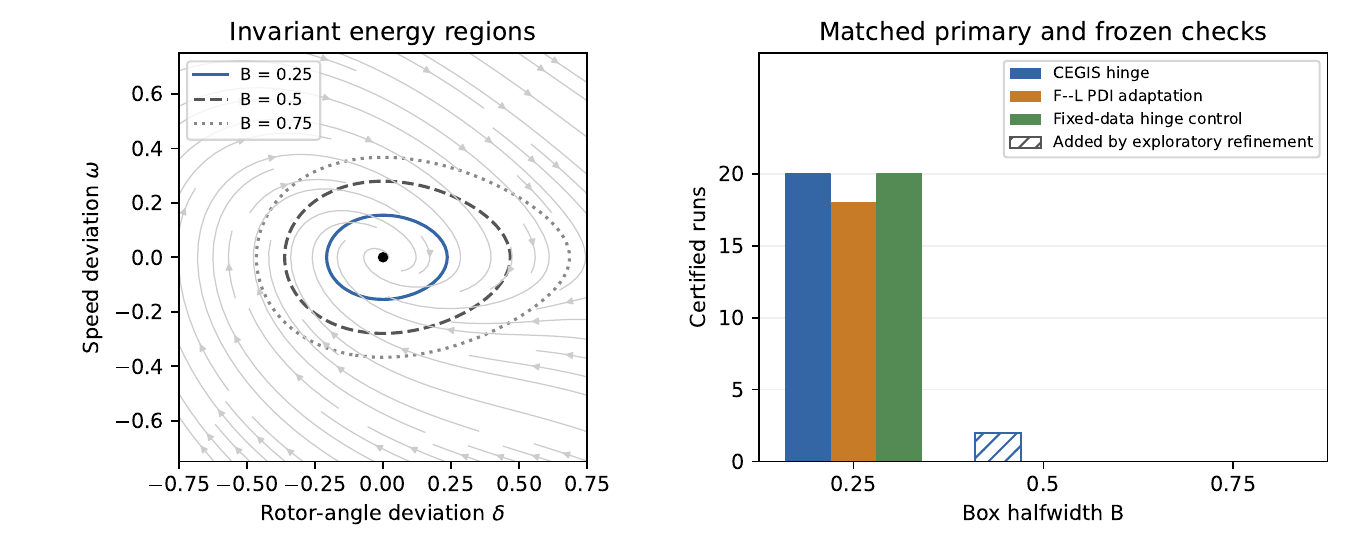}
\caption{Published swing-model application. Left: invariant in-box energy-region boundaries and diagnostic phase-portrait arrows. Invariance follows from the energy argument, not the plotted trajectories; a region receives the requested rate only when its metric certifies. Right: primary certifications under matched caps, with hatched additions from the explicitly exploratory frozen-candidate refinement. Every domain and seed is included. The PDI bars refer to the controlled objective adaptation.}
\label{fig:application}
\end{figure}

\subsection{Certified constants, costs, and scope}

Tables~\ref{tab:certconstants} and \ref{tab:certsummaries} give complete representative certificates and certified-seed distributions. Each representative is one actual candidate; componentwise medians and ranges describe distributions and do not constitute a joint certificate. The polynomial and pendulum rows use the tighter certified trace lower bound in Remark~\ref{rem:eta}; the neural row retains its saved $na$ conversion. The complete per-seed records preserve the relation between the parameters, oracle bound, slack, metric bounds, improved rate, and overshoot. Table~\ref{tab:timings} reports learner and oracle costs on the recorded machine, including evaluation counts and bound computation.

The results establish sound positive certificates within the stated domains and finite computational budgets. They do not establish completeness of the numerical learners. The scale sequence is constructive, the neural and polynomial learners are not a controlled expressiveness comparison, and the dimensional grid extrapolation assumes a fixed measured ratio. The three-dimensional benchmark extends the implementation beyond the plane but does not remove worst-case subdivision growth. The surrogate transfer uses known truth to certify global residuals. Variational and finite-distance conclusions retain the flow conditions of Remark~\ref{rem:finite} for the relevant true or model dynamics.

\FloatBarrier
\begingroup
\makeatletter
\renewenvironment{table}[1][]{%
  \par\addvspace{\intextsep}\noindent
  \begin{minipage}{\linewidth}\def\@captype{table}%
}{%
  \end{minipage}\par\addvspace{\intextsep}%
}
\makeatother
\IfFileExists{revision_tables.tex}{
\begin{table}[t]\centering\caption{Complete representative certificates. The smallest certified seed is used per study. Upper bounds are rounded upward and lower guarantees downward.}\label{tab:certconstants}
\small\begin{tabular}{lrrr}\hline Quantity & Polynomial & Pendulum & Neural\\\hline
Seed & $0$ & $0$ & $0$\\
$N_{\rm ax}$ & $2001$ & $401$ & $8001$\\
$\tau$ & $0.001415$ & $0.003536$ & $3.536 \times 10^{-4}$\\
$L_h$ & $19.71$ & $1.556$ & $54.02$\\
$L_h\tau$ & $0.02787$ & $0.005501$ & $0.0191$\\
$\underline s$ & $0.02155$ & $0.04321$ & $0.009081$\\
$\underline v$ & $0.8687$ & $2.284$ & $0.1999$\\
$\eta$ & $0.01872$ & $0.09872$ & $0.001816$\\
$a$ & $0.1$ & $0.1$ & $0.1$\\
$b$ & $5.934$ & $2.244$ & $5.237$\\
$\lambda'$ & $0.40157832$ & $0.22200502$ & $0.40017341$\\
$\sqrt{b/a}$ & $7.703$ & $4.737$ & $7.237$\\
\hline\end{tabular}\end{table}
\begin{table}[t]\centering\caption{Certified-seed distributions: median [minimum, maximum], rounded to nearest. Each component is summarized separately; these summaries do not form a joint certificate. Full per-seed constants are in the reproduction.}\label{tab:certsummaries}
\scriptsize\begin{tabular}{llrrr}\hline Quantity & & Polynomial & Pendulum & Neural\\\hline
$L_h$ & & $8.59\ [3.84, 26.8]$ & $1.65\ [0.944, 5.69]$ & $413\ [44.3, 825]$\\
$L_h\tau$ & & $0.0121\ [0.00543, 0.038]$ & $0.00583\ [0.00334, 0.0201]$ & $0.146\ [0.0157, 0.292]$\\
$\underline s$ & & $0.0355\ [0.00541, 0.0557]$ & $0.0425\ [0.00311, 0.0965]$ & $0.0342\ [0.00287, 0.101]$\\
$\eta$ & & $0.0461\ [0.00458, 0.0843]$ & $0.0931\ [0.00721, 0.252]$ & $0.00684\ [5.74 \times 10^{-4}, 0.0203]$\\
$b$ & & $4.25\ [2.46, 8.59]$ & $2.31\ [2.09, 2.54]$ & $5.18\ [3.09, 6.75]$\\
$\lambda'$ & & $0.405\ [0.4, 0.414]$ & $0.22\ [0.202, 0.256]$ & $0.401\ [0.4, 0.402]$\\
$\sqrt{b/a}$ & & $6.52\ [4.96, 9.27]$ & $4.8\ [4.57, 5.04]$ & $7.2\ [5.56, 8.22]$\\
\hline\end{tabular}\end{table}
\begin{table}[t]\centering\caption{Median cost per retraining round (excluding initial verification), with ranges in the reproduction. Oracle time includes bound computation; evaluation counts include screens and enclosure operations.}\label{tab:timings}
\small\begin{tabular}{lrrr}\hline Study & Learner (s) & Oracle (s) & Evaluations\\\hline
Polynomial / fixed coarse & $0.00282$ & $0.0999$ & $1.46 \times 10^{4}$\\
Polynomial / coupled coarse & $1.03$ & $0.107$ & $1.46 \times 10^{4}$\\
Polynomial / operating & $0.524$ & $0.487$ & $4 \times 10^{6}$\\
Pendulum & $0.459$ & $0.207$ & $3.22 \times 10^{5}$\\
Neural & $1.41$ & $11.3$ & $6.4 \times 10^{7}$\\
\hline\multicolumn{4}{l}{\footnotesize Apple M3 Max; Python 3.12.7; one BLAS thread.}\\
\multicolumn{4}{l}{\footnotesize Independent producers may overlap; timings include contention.}\\
\hline\end{tabular}\end{table}
\begin{table}[t]\centering\caption{Budgeted grid refinement of frozen candidates. The last two columns compare valid margin bounds on the final grid; the coordinatewise bound does not change the learner.}\label{tab:refinement}
\small\begin{tabular}{rrrrr}\hline Seed & Final $N_{\rm ax}$ & Checks & Euclidean margin & Coordinate margin\\\hline
$0$ & $1921$ & $5$ & $0.0291$ & $0.0273$\\
$1$ & $961$ & $4$ & $0.0393$ & $0.0386$\\
$2$ & $481$ & $3$ & $0.025$ & $0.0222$\\
\hline\end{tabular}\end{table}
\begin{table}[t]\centering\caption{Matched operating-grid ablation and constrained controls. All operating-grid regimes share seeds and budgets. Bounded comparisons cover coupled margins only. The final row uses the coarse grid.}\label{tab:ablation}
\small\begin{tabular}{lrrrrr}\hline Test / rule & $N_{\rm ax}$ & Cert. & Div. & Stall & Fail\\\hline
Raw / fixed & 2001 & 20 & 0 & 0 & 0\\
Raw / coupled & 2001 & 20 & 0 & 0 & 0\\
Normalized / fixed & 2001 & 20 & 0 & 0 & 0\\
Normalized / coupled & 2001 & 19 & 1 & 0 & 0\\
Raw / bounded coupled & 2001 & 20 & 0 & 0 & 0\\
Normalized / bounded coupled & 2001 & 19 & 1 & 0 & 0\\
Normalized / bounded coarse & 121 & 0 & 20 & 0 & 0\\
\hline\end{tabular}\end{table}
\begin{table}[t]\centering\caption{Short-horizon sensitivity on a common coarse diagnostic grid. These counts are not operating-grid success rates. The pendulum row is a separate matched fixed-margin control at its original operating resolution.}\label{tab:sensitivity}
\small\begin{tabular}{lrrrr}\hline Setting & Seeds & Cert. & Div. & Stall\\\hline
$\gamma=1.1$, $q_c=60$ & 5 & 0 & 3 & 2\\
$\gamma=1.3$, $q_c=60$ & 5 & 0 & 5 & 0\\
$\gamma=2.0$, $q_c=60$ & 5 & 0 & 5 & 0\\
$\gamma=1.3$, $q_c=10$ & 5 & 0 & 5 & 0\\
$\gamma=1.3$, $q_c=all$ & 5 & 0 & 5 & 0\\
Pendulum / fixed & 20 & 20 & 0 & 0\\
\hline\end{tabular}\end{table}
\begin{table}[t]\centering\caption{Adaptive verification and fitted-surrogate transfer. Box and grid-operation counts have different units. The surrogate residual bounds use known polynomial truth over the entire box.}\label{tab:extensions}
\small\begin{tabular}{lr}\hline Quantity & Result\\\hline
3D certified / seeds & 11 / 20\\
3D inconclusive & 9\\
3D median retraining rounds & $2$\\
3D median boxes (certified runs) & $1.023 \times 10^{4}$\\
3D representative seed & $0$\\
3D $a$ & $0.1$\\
3D $b$ & $6.591$\\
3D raw $\eta$ & $1 \times 10^{-4}$\\
3D rate $\lambda'$ & $0.40000758$\\
3D overshoot & $8.119$\\
Frozen seed 0 adaptive boxes & $121$\\
Frozen seed 0 grid / adaptive time (s) & $3.19 / 0.116$\\
Frozen seed 1 adaptive boxes & $47$\\
Frozen seed 1 grid / adaptive time (s) & $3.02 / 0.0413$\\
Frozen seed 2 adaptive boxes & $83$\\
Frozen seed 2 grid / adaptive time (s) & $3.17 / 0.0766$\\
Surrogate trajectory samples & $300$\\
Surrogate derivative-fit RMSE & $3.36 \times 10^{-15}$\\
Surrogate $\eta$ & $0.005$\\
Metric $b$ & $4.428$\\
Metric $\mu_1$ & $1.206$\\
Residual $\delta_0$ & $5.012 \times 10^{-4}$\\
Residual $\delta_1$ & $4.673 \times 10^{-4}$\\
Remaining raw margin & $2.584 \times 10^{-4}$\\
True-system rate $\lambda''$ & $0.40002919$\\
Overshoot $\sqrt{b/a}$ & $6.654$\\
Transfer certified & Yes\\
\hline\end{tabular}\end{table}
}{}
\IfFileExists{application_tables.tex}{\begin{table}[t]
\centering\small
\caption{Matched swing-model comparison. Each row includes every seed. Updates and times are medians over all runs, including failures; early certification stops a run. Equal update and oracle caps do not imply equal wall time.}
\label{tab:application}
\begin{tabular}{clrrrr}
\hline
$B$ & Learner & Cert./seeds & Updates & Learn (s) & Verify (s) \\
\hline
$0.25$ & CEGIS hinge & 20/20 & $400$ & $0.347$ & $0.0715$ \\
$0.25$ & F--L PDI adaptation & 18/20 & $400$ & $0.358$ & $0.142$ \\
$0.25$ & Fixed-data hinge control & 20/20 & $400$ & $0.341$ & $0.0629$ \\
$0.5$ & CEGIS hinge & 0/20 & $2000$ & $1.76$ & $2.74$ \\
$0.5$ & F--L PDI adaptation & 0/20 & $2000$ & $1.8$ & $0.00162$ \\
$0.5$ & Fixed-data hinge control & 0/20 & $2000$ & $1.72$ & $0.00162$ \\
$0.75$ & CEGIS hinge & 0/20 & $2000$ & $1.76$ & $0.00158$ \\
$0.75$ & F--L PDI adaptation & 0/20 & $2000$ & $1.8$ & $0.00157$ \\
$0.75$ & Fixed-data hinge control & 0/20 & $2000$ & $1.72$ & $0.00157$ \\
\hline
\end{tabular}
\end{table}

\begin{table}[t]
\centering\small
\caption{Paired certification outcomes on the swing model. Counts compare the CEGIS hinge learner against the indicated comparator using matched seeds. These descriptive counts do not rank the original published tools.}
\label{tab:applicationpaired}
\begin{tabular}{clrrrr}
\hline
$B$ & Comparator & Both & CEGIS only & Comparator only & Neither \\
\hline
$0.25$ & F--L PDI adaptation & 18 & 2 & 0 & 0 \\
$0.25$ & Fixed-data hinge control & 20 & 0 & 0 & 0 \\
$0.5$ & F--L PDI adaptation & 0 & 0 & 0 & 20 \\
$0.5$ & Fixed-data hinge control & 0 & 0 & 0 & 20 \\
$0.75$ & F--L PDI adaptation & 0 & 0 & 0 & 20 \\
$0.75$ & Fixed-data hinge control & 0 & 0 & 0 & 20 \\
\hline
\end{tabular}
\end{table}

\begin{table}[t]
\centering\small
\caption{Exploratory verification-budget refinement of every frozen final candidate. The larger common box budget was selected after preliminary primary outcomes. Additional time and total primary-plus-refinement time are medians over all runs; no learning is repeated and the primary outcomes remain unchanged.}
\label{tab:applicationrefinement}
\begin{tabular}{clrrrr}
\hline
$B$ & Learner & Cert./seeds & Additional cert. & Extra (s) & Total (s) \\
\hline
$0.25$ & CEGIS hinge & 20/20 & 0 & $0.0701$ & $0.488$ \\
$0.25$ & F--L PDI adaptation & 18/20 & 0 & $0.14$ & $0.67$ \\
$0.25$ & Fixed-data hinge control & 20/20 & 0 & $0.0613$ & $0.469$ \\
$0.5$ & CEGIS hinge & 2/20 & 2 & $7.01$ & $11.4$ \\
$0.5$ & F--L PDI adaptation & 0/20 & 0 & $2.81 \times 10^{-4}$ & $1.8$ \\
$0.5$ & Fixed-data hinge control & 0/20 & 0 & $2.52 \times 10^{-4}$ & $1.72$ \\
$0.75$ & CEGIS hinge & 0/20 & 0 & $2.39 \times 10^{-4}$ & $1.77$ \\
$0.75$ & F--L PDI adaptation & 0/20 & 0 & $2.44 \times 10^{-4}$ & $1.8$ \\
$0.75$ & Fixed-data hinge control & 0/20 & 0 & $2.43 \times 10^{-4}$ & $1.72$ \\
\hline
\end{tabular}
\end{table}

\begin{table}[t]
\centering\small
\caption{Post hoc analytical constant-metric baseline. A single fixed matrix is checked on each whole box at the shared rate and strict margin. There are no training seeds or timing comparisons. A positive diagonal witness rules out this matrix on that box, not other metrics or contraction on its interior energy subset.}
\label{tab:applicationconstant}
\begin{tabular}{clc}
\hline
$B$ & Fixed-$P$ outcome & Proved lower bound on $[C_P(k(B))]_{11}$ \\
\hline
$0.25$ & Certified on whole box & --- \\
$0.5$ & Violates requested rate & $0.0764$ \\
$0.75$ & Violates requested rate & $0.3244$ \\
\hline
\end{tabular}
\end{table}
}{}
\endgroup

\FloatBarrier
\section{Conclusion}
\label{sec:conclusion}

We studied counterexample-guided synthesis of contraction metrics with sound verification. Computable polynomial constants and outward-rounded checks establish the meaning of a positive answer. An idealized finite-grid result separates sufficient feasibility and boundedness assumptions from the behavior of finite-budget Adam. Scaling and fixed-margin stationary updates explain why an already valid metric can remain uncertified; normalization and margin coupling address the stated mechanisms without guaranteeing convergence. Verified slack improves the differential contraction rate and supports explicit transfer under bounded model error.

The matched polynomial operating-grid ablation did not demonstrate an advantage in certification counts from trace normalization or margin coupling within the tested budgets. The contribution centers on the analysis of failure mechanisms and sound certification; these results do not establish the training choices as necessary or generally superior. The experiments also distinguish a constructive scale sequence, adaptive verification, and a known-truth fitted-surrogate test. A published power-system swing model adds a matched neural PDI-loss comparison and a fixed-data hinge control, whose primary counts match CEGIS. Verified analytical constant metrics certify both the polynomial benchmark and the smallest swing-model box, so those successes do not establish a need for neural state dependence. Exploratory refinement certifies additional frozen CEGIS candidates at a larger verification budget. Successful certificates support synchronization guarantees on explicit invariant energy subsets; uncertified boxes and the controlled adaptation's scope limit broader conclusions. The next methodological questions concern learner progress under practical constraints and verification cost on less structured higher-dimensional systems. Extensions to controller synthesis could use dual control-contraction-metric formulations \cite{manchester2017} or direct joint controller-and-primal-metric learning \cite{davydov2025partitioning,harapanahalli2026}.

\subsection*{Data and code availability}
Release \texttt{v1.2.2} contains the manuscript without reviewer line numbers, tested arXiv sources, complete reproduction, and independent-review package \cite{nsp_release}. Its fresh fifteen-stage run regenerates all experiments, verifies both constant-metric baselines, and rebuilds the paper. The archive records the fixed protocol, every seed outcome and positive proof cover, exact and directed baseline checks, hashes, numerical environment, and logs. This release clarifies descriptive rounding, the polynomial metric choice, and its proof margin. The immutable \texttt{v1.2.1} release remains unchanged. The cited GitHub record confirms publication and asset immutability.

\subsection*{Acknowledgment of AI assistance}
OpenAI Codex assisted with mathematical analysis and exposition, literature searches and reference checking, manuscript drafting and revision, software implementation, and execution and checking of the computational pipeline. Anthropic Claude supplied a critical review that informed the revision. The author retains responsibility for the final manuscript, code, interpretation of the results, and scientific claims.

\appendices
\section{Polynomial coefficient bounds and proof of Lemma~\ref{lem:certconst}}
\label{app:coeff}
\begin{lemma}[Coefficient bound]
\label{lem:coeff}
Let $p(x) = \sum_{a} c_a\, x^a$ be a polynomial on $\X = [-B, B]^n$, with multi-index $a$ and $|a| = \sum_i a_i$. Then
\[
\sup_{x \in \X} |p(x)| \;\le\; \sum_{a} |c_a|\, B^{|a|} .
\]
Subdividing $\X$ into congruent subboxes and recentering $p$ on each by the exact binomial shift $x \mapsto c + y$ yields the same bound applied at the subbox half-width, and the resulting refined bound is monotone nonincreasing in the subdivision depth.
\end{lemma}

\begin{proof}
On $\X$ each monomial satisfies $|x^a| \le B^{|a|}$, and the triangle inequality gives the first claim. The recentered polynomial equals $p$ on its subbox, so the bound applied at the half-width is valid there, and the maximum over subboxes bounds $\sup_\X |p|$. Refining a subbox can only lower its local bound, which gives monotonicity.
\end{proof}

\begin{proof}[Proof of Lemma~\ref{lem:certconst}]
Weyl's inequality bounds $|g(x)-g(y)|$ by $\|C(x)-C(y)\|_F$. Integrating along the segment in the convex box yields $\|C(x)-C(y)\|_F\le\sum_kD_k|x_k-y_k|$, and Cauchy--Schwarz gives (i). For (ii), $v=n\alpha+\|L_\theta\|_F^2\ge n\alpha$ and $|v(x)-v(y)|\le L_v\|x-y\|_2$. The quotient identity
\[
\tilde g(x)-\tilde g(y)=\frac{g(x)-g(y)}{v(x)}+\tilde g(y)\frac{v(y)-v(x)}{v(x)}
\]
gives the claimed normalized bound. The local estimate $|\tilde g|\le\|C\|_F/v$ gives $\widehat G$. For (iii), $M_\theta\succeq\alpha I$ and $\lambda_{\max}(M_\theta)\le\|M_\theta\|_F$; apply entrywise coefficient bounds. Finally $\|\partial_k M_\theta\|_2\le\|\partial_k M_\theta\|_F$, and summing the squared entrywise bounds gives (iv). The same quotient identity coordinate by coordinate proves Remark~\ref{rem:coordinate}.
\end{proof}

\section{Directed-rounded verification}
\label{app:numerics}
The verifier interprets the stored binary64 parameters as exact real inputs. It constructs polynomial coefficients, derivatives, and shifts with exact rational arithmetic, then converts bounds outward. Interval operations enclose each elementary binary64 operation with adjacent representable endpoints. Transcendental enclosures use directed MPFR evaluation through gmpy2 \cite{fousse2007}; large batches use a tabulated interpolant with a rigorous derivative-based remainder. Linear interpolation on cells of width $h$ has error at most $h^2\sup|u^{\prime\prime}|/8$, so directed nodal enclosures and a bound on the second derivative enclose the interpolant error. The verifier bounds the covering radius from the largest gap between the actual stored grid nodes. Ordinary floating-point evaluations can reject a candidate, but a positive answer requires an outward-rounded upper bound at every grid point to clear the outward-rounded product $L_h\tau$. The neural verifier may enclose groups of grid nodes by interval boxes, subdividing unresolved groups down to individual nodes; this changes the cost of establishing the same grid inequalities. Returned training points are diagnostic candidates; a screening rejection or inconclusive interval comparison does not prove a contraction violation. The returned certified slack is rounded downward, as are sufficient robustness budgets. Overflow and invalid interval operations fail verification. These guarantees rely on IEEE binary64 arithmetic with round-to-nearest and gradual underflow and on MPFR's directed-rounding contract; the reproduction checks test these arithmetic assumptions. Finite-difference and dense-sampling diagnostics remain implementation checks rather than substitutes for the enclosure argument.

\end{document}